\documentclass[universe,review,accept,moreauthors,pdftex]{Definitions/mdpi}

\usepackage{dcolumn}
\usepackage{bm}
\usepackage{ifpdf}
\usepackage{hyperref}
\usepackage{bm}
\usepackage{xcolor,color,graphicx,graphics}
\usepackage[portuguese, english]{babel}
\usepackage[OT1]{fontenc}
\usepackage{latexsym,amssymb,amsmath,amsfonts}
\usepackage{makeidx}
\usepackage{epsfig}
\usepackage{natbib}
\usepackage{epstopdf}
\usepackage{mathrsfs}
\usepackage{enumerate}
\usepackage{tikz}

\newcommand{\iprod}{\mathbin{\lrcorner}}
\newcommand*{\rom}[1]{\expandafter\@slowromancap\romannumeral #1@}

\newcommand{\be}{\begin{equation}}
  \newcommand{\ee}{\end{equation}}
\newcommand{\ben}{\begin{eqnarray*}}
  \newcommand{\een}{\end{eqnarray*}}
\newcommand{\bea}{\begin{eqnarray}}
  \newcommand{\eea}{\end{eqnarray}}
\newcommand{\bdm}{\begin{displaymath}}
  \newcommand{\edm}{\end{displaymath}}
\newcommand{\ba}{\begin{align}}
  \newcommand{\ea}{\end{align}}

\firstpage{1}
\pubvolume{6}
\issuenum{1}
\articlenumber{238}
\pubyear{2020}
\copyrightyear{2020}
\history{Received: 13 November 2020; Accepted: 7 December 2020; Published: 11 December 2020}

\Title{Fundamental Symmetries and Spacetime Geometries in Gauge Theories of Gravity---Prospects for Unified Field Theories} 

\newcommand\orcidCabral{{\href{https://orcid.org/0000-0003-2124-5894}{\orcidicon}}}
\newcommand\orcidLobo{{\href{https://orcid.org/0000-0002-9388-8373}{\orcidicon}}}
\newcommand\orcidDiego{{\href{https://orcid.org/0000-0003-3984-9864}{\orcidicon}}}

\Author{Francisco Cabral\orcidCabral 
 $^{1,}$*, Francisco S. N. Lobo\orcidLobo 
 $^{1}$ and Diego Rubiera-Garcia\orcidDiego $^{2}$}

\AuthorNames{Firstname Lastname, Firstname Lastname and Firstname Lastname}

\address{%
$^{1}$ \quad Instituto de Astrof\'{\i}sica e Ci\^{e}ncias do Espa\c{c}o, Faculdade de
Ci\^encias da Universidade de Lisboa, Edif\'{\i}cio C8, Campo Grande,
P-1749-016 Lisbon, Portugal; fslobo@fc.ul.pt\\
$^{2}$ \quad Departamento de F\'isica Te\'orica and IPARCOS, Universidad Complutense de Madrid, \mbox{E-28040
Madrid, Spain;} drubiera@ucm.es}

\corres{Correspondence: ftcabral@fc.ul.pt}

\abstract{Gravity can be formulated as a gauge theory by combining symmetry principles and geometrical methods in a consistent mathematical framework. The gauge approach to gravity leads directly to non-Euclidean, post-Riemannian spacetime geometries, providing the adequate formalism for metric-affine theories of gravity with curvature, torsion and non-metricity. In~this paper, we analyze the structure of gauge theories of gravity and consider the relation between fundamental geometrical objects and symmetry principles as well as different spacetime paradigms. Special attention is given to Poincar\'{e} gauge theories of gravity, their field equations and  Noether conserved currents, which are the sources of gravity. We then discuss several topics of the gauge approach to gravitational phenomena, namely, quadratic Poincar\'{e} gauge models, the~Einstein-Cartan-Sciama-Kibble theory, the teleparallel equivalent of general relativity, quadratic metric-affine Lagrangians, non-Lorentzian connections, and the breaking of Lorentz invariance in the presence of non-metricity.  We also highlight the probing of post-Riemannian geometries with test matter. Finally, we briefly discuss some perspectives regarding the role of both geometrical methods and symmetry principles towards unified field theories and a new spacetime paradigm, motivated from the gauge approach to gravity.}


\begin{document}
\section{Introduction}

The success of Einstein's General Theory of Relativity (GR) to describe the behaviour of the gravitational interaction continues to amaze us. It has passed all tests performed so far: Solar System observations and binary pulsars \cite{Will:2014kxa}, stellar orbits around the central galactic black hole \cite{Hees:2017aal}, gravitational waves (GWs) from coalescing compact objects (black holes and neutron stars) \cite{Abbott:2016blz,TheLIGOScientific:2017qsa,TheLIGOScientific:2016src,Abbott:2017oio}, or the indirect observation of the black hole horizon with the Event Horizon Telescope \cite{Akiyama:2019cqa}, among others. At the same time, it provides us with the observationally valid framework for the standard cosmological paradigm when supplemented with the (cold) dark matter and dark energy hypothesis \cite{Bull:2015stt}.

Soon after Einstein formulated GR in its final form, Weyl introduced the notion of gauge transformations in an attempt to unify gravity and electromagnetism \cite{Weyl:1919fi}. By extending the local Lorentz group to include scale transformations (dilatations), he was led to assume what we call nowadays a Riemann-Weyl spacetime geometry, namely, a post-Riemann geometry with (the trace-vector part of) non-metricity in addition to the familiar curvature of GR. Later on, it was discovered that the electromagnetic field itself is intimately related to local internal symmetries, under the $U(1)$ group that acts on the four-spinor fields of charged matter \cite{Weyl2}. In the 1950s, Yang and  Mills \cite{Yang:1954ek} further explored the notion of gauge symmetries in field theories going beyond the $U(1)$ group to include non-abelian Lie groups ($SU(2)$),  in order to address nuclear physics, while Utiyama \cite{Utiyama:1956sy} extended the gauge principle to all semi-simple Lie groups including the Lorentz group.

The gauge principle is based on the localization of the  rigid global symmetry group of a field theory, introducing a new interaction described by the gauge potential. The latter is a compensating field that makes it possible for the matter Lagrangian to be locally invariant under the symmetry group and is included in the covariant derivative of the theory. There is a clear geometrical interpretation of the gauge potential as the connection of the fiber bundle, which is the manifold obtained from the base spacetime manifold and the set of all fibers. These are attached at each spacetime point and are the (vector, tensor or spinor) spaces of representation of the local symmetries. In the geometrical interpretation, the imposition of local symmetries implies that the geometry of the fiber bundle is non-Euclidean, and the gauge field strengths are the curvatures of such a manifold.

The gauge formulation of gravity was resumed through the works of Sciama \cite{Sciama:1964wt} and Kibble \cite{Kibble:1961ba}, who gauged the (rigid) Poincar\'{e} group of Minkowski spacetime symmetries. They~arrived at what is now known as a Riemann-Cartan (RC) geometry, and to the corresponding Einstein-Cartan-Sciama-Kibble (ECSK) gravity, with non-vanishing torsion and curvature. This is a natural extension of GR, which is able to successfully incorporate the intrinsic spin of fermions as a source of gravity, while passing all weak-field limit tests. Moreover, the theory has no free parameters but rather a single new scale given by Cartan's density, which yields many relevant applications in cosmology and astrophysics \cite{Boehmer:2008ah,Poplawski:2011jz,Poplawski:2012ab,Vakili:2013fra,Bronnikov:2016xvj,Ivanov:2016xjm,Cembranos:2016xqx,Cembranos:2017pcs,Unger:2018oqo}. It is worth noting that ECSK is the simplest of all Poincar\'{e} gauge theories of gravity (PGTG). Beyond the Poincar\'{e} group we have, for instance, the Weyl and the conformal groups, which live on a subset of a general metric-affine geometry, with non-vanishing curvature, torsion and non-metricity (for a detailed analysis and reviews on several topics of the gauge approach to gravity see the remarkable works in References \cite{Blagojevic:2012bc,Blagojevic:2002du,Obukhov}). By extending the gauge symmetry group of gravity, one is naturally led to extend the spacetime geometry paradigm as well.  The geometrical methods and symmetry principles, motivated by gauge theories of gravity, together with post-Riemannian geometries and Yang-Mills gauge fields, are expected to play a key role towards a unified  field theory. The corresponding physical description might lead to a new spacetime paradigm extending the notion of the classical spacetime manifold where physical fields propagate on it and affect its geometry. Instead, it is plausible to expect a consistent unified physical manifold, where the properties of spacetime, matter fields and vacuum are manifestations of the same (unified) fundamental physical reality.

The main aim of this work is to review and discuss the basics and recent developments within gauge theories of gravity and post-Riemann geometries in connection to their  perspectives for achieving an unified field theory. This is a vital part of the effort to understand the nature of spacetime and gravity in those regimes where the standard picture provided by GR may break down, challenging our current ideas on the spacetime paradigm. This article complements our previous works:  an extension of the Einstein-Cartan-Dirac theory where an electromagnetic (Maxwell) contribution minimally coupled to torsion is introduced (which breaks the $U(1)$ gauge symmetry \cite{Cabral:2019gzh}), and its extension to analyze in detail the physics of this model with  Dirac and Maxwell fields minimally  coupled to the spacetime torsion  \cite{Cabral:2020mst,Cabral:2020mzw}.

This work is organized as follows: in Section \ref{sec:II} we introduce the fundamental geometrical objects of the spacetime manifold in the formalism of exterior forms and its relation to spacetime symmetries. In Section \ref{sec:III} we briefly summarize the metric-affine approach to gravity using these geometrical methods and symmetry principles, and discuss several paradigms for the spacetime geometry, including the general metric-affine geometry. In Section \ref{sec:IV}, we outline the general structure of the gauge approach to gravity. We illustrate it with the case of PGTG, consider the field equations, the Noether conserved currents (sources of gravity), and include discussions on several topics in gauge theories of gravity, such as quadratic Poincar\'{e} gauge models, the ECSK theory, the teleparallel equivalent of GR (an example of a translational gauge model), quadratic metric-affine Lagrangians, the breaking of Lorentz invariance in the presence of non-metricity, the nature of the hypermomentum currents and the probing of post-Riemann geometries with test matter. Finally, in Section \ref{Conclusion} we discuss some perspectives for unified field theories and a new spacetime paradigm motivated from the geometrical methods and symmetry principles of the gauge approach to gravity.

\section{Spacetime Symmetries and Post-Riemann Geometries\label{sec:II}}
\unskip

\subsection{Fundamental Geometrical Structures of Spacetime}
Let us begin by considering a four-dimensional differential manifold $\mathcal{M}$ as an approximate representation of the physical spacetime, and introduce the fundamental geometrical objects and its relation to the group theory of spacetime symmetries. At each point $P$ of the spacetime manifold we introduce the set of four linearly independent vectors $\lbrace\bar{e}_{0},\bar{e}_{1},\bar{e}_{2},\bar{e}_{3}\rbrace$, with $\bar{e}_{0}\equiv \partial_{0}$, $\bar{e}_{1}\equiv \partial_{1}$, $\bar{e}_{2}\equiv \partial_{2}$, $\bar{e}_{3}\equiv \partial_{3}$, which constitute the vector coordinate (holonomic) basis, where each vector is tangent to a coordinate line. This is called a linear frame basis. Similarly, at the same point we introduce the dual co-frame basis $\lbrace\bar{\theta}^{0},\bar{\theta}^{1},\bar{\theta}^{2},\bar{\theta}^{3}\rbrace$, with $\bar{\theta}^{0}\equiv dx^{0}$, $\bar{\theta}^{1}\equiv dx^{1}$, $\bar{\theta}^{2}\equiv dx^{2}$, $\bar{\theta}^{3}\equiv dx^{3}$, which satisfies $\bar{e}_{b}\iprod\bar{\theta}^{a}=\delta^{a}_{b}$\footnote{The symbol $\iprod$ stands for the interior product defined in differential topology and linear algebra, also called contraction operator, which gives a contraction between a $p$-form and a vector, resulting in a $(p-1)$-form.}. Any of these basis can be called the natural (coordinate/holonomic) frame/co-frame. This geometrical structure comes naturally with the notion of coordinates on the spacetime manifold, as it is intrinsic to the coordinates structure.

Note that one can choose any set of linearly independent vectors/co-vectors to form arbitrary linear frames and co-frames. To this effect, we consider the independent combinations $e_{b}=e_{b}^{\;\;\nu}\partial_{\nu}$ and $\theta^{a}=\theta^{a}_{\;\;\mu}dx^{\mu}$. The indices $a,b=0,1,2,3$ are called anholonomic indices, sometimes known as symmetry or group indices, and play a fundamental role in the gauge approach to gravity due to its connection to spacetime symmetries. It is clear that for the natural frame/co-frame we have $\bar{e}_{b}=\delta_{b}^{\;\;\nu}\partial_{\nu}$ and $\bar{\theta}^{a}=\delta^{a}_{\;\;\mu}dx^{\mu}$. For arbitrary (non-coordinate) anholonomic vector basis, the Lie brackets is non-vanishing,  $\left[U,V\right]\equiv\pounds_{U}V \neq 0$, for any two vectors $U, V$ in the basis. In relation to this one can show, using the definitions and duality relations already introduced, the following algebra \mbox{$\left[e_{a},e_{b}\right]=f_{ab}{}^{c}e_{c}$},
where the objects $f_{ab}{}^{c}$ are typically known as the (group) structure constants. This~algebra can be used to characterize the local spacetime symmetries of the tangent/cotangent~spaces.

In four dimensions, the set of vector valued 1-forms $\theta^{a}$ constitute 16 independent components $\theta^{a}{}_{\mu}$ (the tetrads) and are the potentials for the group of (local) spacetime translations $T(4)$\footnote{This group is actually related to diffeomorphisms  (and to our freedom to choose any system of coordinates without altering the physical description, no matter the theory of the gravitational field we are working with), playing the role of gauge transformations of gravity \cite{Obukhov:2020uan}.}.  This group has four generators, which therefore entails four potentials and four field strengths $T^{a}=\tfrac{1}{2}T^{a}{}_{\mu\nu}dx^{\mu}\wedge dx^{\nu}$, the latter being a vector valued 2-form field, and corresponds to the torsion of the spacetime manifold. It is given by

\begin{equation}
T^{a}=D\theta^{a}=d\theta^{a}+\Gamma^{a}_{\;\;b}\wedge\theta^{b} \ ,
\end{equation}
where $D$ is the (gauge) covariant exterior derivative, $d$ is the exterior derivative, a kind of curl operator that raises the degree of any $p$-form, $\wedge$ is the wedge product\footnote{Accordingly, $d\theta^{a}$ is a 2-form given by $d\theta^{a}=\partial_{[\mu}\theta^{a}_{\;\;\nu]}dx^{\mu}\wedge dx^{\nu}$. Similarly, if $v$ is a $p$ form and $w$ is a $k$-form, then $v\wedge w$ is a $(p+k)$-form with $v\wedge w=(-1)^{p\times k}w\wedge v$. The (gauge) covariant exterior derivative of a generic tensor valued $p$-form, denoted by $V^{a}_{\;\;b}$, used in this paper is given by $DV^{a}_{\;\;b}=dV^{a}_{\;\;b}+\Gamma^{a}_{\;\;c}\wedge V^{c}_{\;\;b}+(-1)^{p}\Gamma^{c}_{\;\;b}\wedge V^{a}_{\;\;c}$.} and in the second term on the right-hand side, sometimes called the non-trivial part, is the linear connection 1-form $\Gamma^{a}_{\;\;b}$.
The torsion 2-form has $4\times 6=24$ independent components given explicitly by

\begin{equation}
T^{a}_{\;\;\mu\nu}=2\partial_{[\mu}\theta^{a}_{\;\;\nu]}+2\Gamma^{a}_{\;\;c[\mu}\theta^{c}_{\;\;\nu]} \,
\end{equation}
where brackets denote antisymmetrization.

Let us now turn our attention to the linear (affine) connection  $\Gamma^{a}{}_{b}$, which is a tensor valued 1-form that connects neighbouring  points of the manifold. Accordingly, if $v=v^{a}e_{a}$ is a vector parallel-transported by an infinitesimal displacement $\delta x$, the difference between the parallely-displaced vector and the original vector is given by

\begin{equation}
v_{\parallel}^{a}(x+\delta x)-v^{a}(x)=\delta_{\parallel}v^{a}=-\Gamma^{a}_{\;\;b}v^{b},
\end{equation}
with $\Gamma^{a}_{\;\;b}=\Gamma^{a}_{\;\;b\mu}dx^{\mu}$.
Therefore, upon an arbitrary infinitesimal displacement from a given point in the spacetime manifold, the linear frame is transformed (e.g., under a Lorentz rotation or some more generic linear transformation) and the connection gives a measure of such a change in the linear frame. It is worth emphasizing that the introduction of the affine structure of the spacetime manifold $\mathcal{M}$ is mandatory in order to be able to compare  between tensor quantities on different spacetime points. Indeed, the affine connection allows for the definition of a covariant derivative, and in this way it establishes a rule for the parallel transport of tensors along curves of $\mathcal{M}$. Moreover, it  allows  to determine the affine geodesics (straightest lines) on $\mathcal{M}$, which do not necessarily coincide with extremal geodesics (``shortest'' paths).

The affine connection is mathematically a very rich object. Among its many properties we underline the following two: (i) the difference of two connections is a tensor; (ii) under a transformation $\Gamma\rightarrow \Gamma+{\bf \tau}$, where ${\bf \tau}$ is a tensor, the covariant derivative of tensors retains its covariance. This means that the components  of the covariant derivative of a tensor still transform as a tensor, which implies that one can incorporate new degrees of freedom in the geometrical (affine) structure of spacetime while preserving the covariance of the equations. Such extensions of the connection involve an extended spacetime geometry, while the inclusion of extra (gauge) degrees of freedom in a field theory requires extending the local symmetry group.  These two facts are inevitably interconnected in gauge theories of gravity.

In four dimensions, the set of tensor valued 1-forms $ \Gamma^{a}_{\;\;b}$ constitute 64 independent components $\Gamma^{a}_{\;\;b\mu}$ and are the potentials for the four-dimensional group of general linear transformations $GL(4,\Re)$. With the definition of the linear frame/coframe the arbitrary (general) non-degenerate linear transformations of spacetime coordinates can be defined and $\Gamma^{a}_{\;\;b}$ turn out to be the generators of such a group of transformations. It has 16 generators, therefore we have 16 potentials which are analogous to the Yang-Mills potentials of the $SU(3)$ group. The associated field strength $R^{a}_{\;\;b}$ is a tensor valued 2-form field $R^{a}_{\;\;b}=\tfrac{1}{2}R^{a}_{\;\;b\mu\nu}dx^{\mu}\wedge dx^{\nu}$, corresponding to the curvature of the spacetime manifold, which can be written explicitly as

\begin{equation}
R^{a}_{\;\;b}=d\Gamma^{a}_{\;\;b}+\Gamma^{a}_{\;\;c}\wedge \Gamma^{c}_{\;\;b} \ .
\end{equation}

The curvature 2-form has $16\times 6=96$ independent components, given by

\begin{equation}
R^{a}_{\;\;b\mu\nu}=2\partial_{[\mu}\Gamma^{a}_{\;\;b\vert\nu]}+2\Gamma^{a}_{\;\;c[\mu}\Gamma^{c}_{\;\;b\vert\nu]} \ .
\end{equation}

The linear connection 1-form can be decomposed according to

\begin{equation}
\label{independentconnection}
\Gamma_{ab}=\tilde{\Gamma}_{ab}+N_{ab}=\tilde{\Gamma}_{ab}+\dfrac{1}{2}Q_{ab}+N_{[ab]} \ ,
\end{equation}
where the Levi-Civita part of the connection, $\tilde{\Gamma}_{ab}$, obeys the Cartan structure equation $d\theta^{a}+\tilde{\Gamma}^{a}_{\;\;b}\wedge\theta^{b}=0$, and $N^{a}_{\;\;b}$ is the so-called distortion 1-form characterizing the post-Riemannian geometries. In particular, one finds that $Q_{ab}=2N_{(ab)}$ and $T^{a}=N^{a}_{\;\;b}\wedge \theta^{b}$. If the linear connection obeys the condition $\Gamma^{ab}=-\Gamma^{ba}$, then it is called a Lorentzian connection (or spin connection) and corresponds to 24 independent components. This is the case when the linear connection is the potential for the Lorentz group $SO(1,3)$ (not the full linear group GL(4,$\Re$)). Accordingly, as we shall see later, the Lorentzian connection is the linear connection of PGTG.

Though one can argue that the spacetime metric is not as fundamental as the affine connection (and indeed a purely affine formulation of gravity is possible, see for example, Reference \cite{Poplawski:2007rx}), one~can introduce it in order to measure time and space intervals as well as angles. Thus, let~us define the Lorentzian metric as the $(0,2)$ tensor $g=g_{ab}\theta^{a}\otimes\theta^{b}$, where $a,b=0,1,2,3$ are anholonomic indices, and the spacetime metric components in the coordinate frame are given by $g_{\mu\nu}=\theta^{a}_{\;\;\mu}\theta^{a}_{\;\;\nu}g_{ab}$.
The~Lorentzian metric $g_{ab}$ is the metric of the tangent space required to compute the inner product between anholonomic vectors and make a map between these vectors and the corresponding dual co-vectors $v_{b}=g_{ab}v^{a}$. As for the spacetime metric $g_{\mu\nu}$, it establishes maps between the contravariant components of vectors and the covariant components of the corresponding dual covectors in the coordinate (holonomic basis), and defines the inner products $g(u,v)=g_{\alpha\beta}u^{\alpha}v^{\beta}=u_{\alpha}v^{\beta}$.
Therefore, the spacetime metric can be seen as the deformation of the Lorentzian (tangent space) metric according to $g_{\mu\nu}=\Omega^{ab}_{\;\;\;\;\mu\nu}g_{ab}$, with   $\Omega^{ab}_{\;\;\;\;\mu\nu}\equiv\theta^{a}_{\;\;\mu}\theta^{a}_{\;\;\nu}$ being a deformation tensor. Since the tangent space has a pseudo-Euclidean geometry, then if the spacetime manifold has non-Euclidean geometry the deformation tensor has to vary from point to point. From now on we consider geometries where the spacetime metric is symmetric $g_{\alpha\beta}=g_{\beta\alpha}$, non-degenerate, $g={\rm det}(g_{\mu\nu})\neq 0$, and~determines the local line element $ds^{2}=g_{\alpha\beta}dx^{\alpha}dx^{\beta}$ with a Lorentzian signature ($\pm2$). As for the metric $g_{ab}$, it is possible to see it as a kind of potential whose corresponding field strength, $Q_{ab}=Dg_{ab}=dg_{ab}+\Gamma^{c}_{\;a}\wedge g_{cb}+\Gamma^{c}_{\;b}\wedge g_{ac}=2\Gamma_{(ab)}$, is a tensor valued 1-form $Q_{ab}=Q_{ab\mu}dx^{\mu}$ with $10\times 4=40$ independent components $Q_{ab\mu}$. This field strength corresponds to the non-metricity tensor valued 1-form and one concludes that, if the connection is non-Lorentzian ($\Gamma^{ab}\neq-\Gamma^{ba}$), then~non-metricity is non-vanishing.

In (pseudo) Euclidean geometries such as that of  Minkowski spacetime, there is always a coordinate system where the components of the connection (the Christoffel symbols) and their derivatives vanish, whereas in a (pseudo) Riemann geometry with non-vanishing curvature, one~can find a local geodesic system of coordinates where the connection vanishes and the metric is given by the Minkowski metric (a freely falling frame), but the derivatives of the connection cannot be set to zero. In such geometries the so-called Levi-Civita connection and the metric are fundamentally related and thus not independent from each other. Indeed, in such a case the metricity condition (vanishing of the covariant derivatives of the metric) holds, implying that the connection is proportional to the first derivatives of the metric (recall that the Levi-Civita connection is the only symmetric connection that obeys the metricity condition) and as such, in GR, the presence of a physical gravitational field is traced to the non-vanishing of the second derivatives of the metric. The Weyl part of the Riemannian curvature is not absent in a freely falling frame, which translates into tidal effects. But both the symmetry of a connection and the metricity condition can be relaxed leading to more general geometries with torsion and non-metricity respectively, as we shall see next.

\subsection{The Decomposition of the Affine Connection} \label{eq:section2b}

We are now ready to bring here the well known result that any affine connection can be decomposed into three independent components. In holonomic (coordinate) components they are expressed as

\begin{equation} \label{eq:affcons}
\Gamma^{\lambda}{}_{\mu\nu} =
	\tilde{\Gamma}^{\lambda}{}_{\mu\nu} +
	K^{\lambda}{}_{\mu\nu}+
	L^{\lambda}{}_{\mu\nu}\,,
\end{equation}
where $\tilde{\Gamma}^{\lambda}{}_{\mu\nu}$ is the Levi-Civita connection associated to the Riemannian curvature
\begin{equation}
{\tilde{R}^\alpha}_{\beta\mu\nu}=2\partial_{[\mu}{\tilde{\Gamma}^\alpha}_{\nu]\beta} +2 {\tilde{\Gamma}^\alpha}_{[\mu \vert \lambda \vert}{\tilde{\Gamma}^\lambda}_{\nu ] \beta} \ ,
\end{equation}
the term $K^{\lambda}{}_{\mu\nu}$ is associated to torsion, $T_{\alpha\beta}^{\lambda}\equiv\Gamma_{[\alpha\beta]}^{\lambda}$, and is denoted contortion:
\begin{equation} \label{eq:contor}
K^{\lambda}{}_{\mu\nu} \equiv  T^{\lambda}{}_{\mu\nu}-2T_{(\mu}{}^{\lambda}{}_{\nu)},
\end{equation}
while the term $L^{\lambda}{}_{\mu\nu}$ is associated to non-metricity $Q_{\rho \mu \nu} \equiv \nabla_{\rho} g_{\mu\nu}$, and is called disformation:
\begin{equation}
L^{\lambda}{}_{\mu\nu} \equiv \frac{1}{2} g^{\lambda \beta} \left( -Q_{\mu \beta\nu}-Q_{\nu \beta\mu}+Q_{\beta \mu \nu} \right).
\end{equation}

This decomposition naturally highlights the metric-affine character of the most general spacetime geometry, made up of curvature, torsion and non-metricity. Let us briefly analyze each component~separately.

\subsubsection{Curvature}

In an holonomic basis the curvature of a connection has the 96 independent components

\begin{equation}
R^{\alpha}_{\;\beta\mu\nu}=\partial_{\mu}\Gamma^{\alpha}_{\;\beta\nu}-\partial_{\nu}\Gamma^{\alpha}_{\;\beta\mu}+\Gamma^{\alpha}_{\;\lambda\mu}\Gamma^{\lambda}_{\;\beta\nu}-\Gamma^{\alpha}_{\;\lambda\nu}\Gamma^{\lambda}_{\;\beta\mu} \ .
\end{equation}

Consider at some point $P$ of the manifold the vectors $U=d/d\lambda$ and $V=d/d\sigma$, tangent to two curves intersecting at $P$, where $\lambda$ and $\sigma$ are some respective affine parameters labelling each curve. Since the curvature tensor  is of $(1,3)$ type, it can be applied (contracted) to $U$ and $V$ and then to some other vector field $Z$, which yields the resulting vector field

\begin{equation}
R(U,V)Z=\nabla_{U}\nabla_{V}Z-\nabla_{V}\nabla_{U}Z-\nabla_{[U,V]}Z \ ,
\end{equation}
where $[U,V]=\pounds_{U}V$ represents the Lie derivative (of $V$ with respect to $U$). To clarify its geometrical meaning, let us consider an infinitesimal closed loop, with $ds^{\mu\nu}$ being the surface element spanned by such a loop. After a parallel transport of some vector $v$ along the loop, it is found that the initial and final vectors do not coincide: there is a rotation with a difference vector

\begin{equation}
\delta v^{\alpha}\approx R^{\alpha}_{\;\beta\mu\nu}v^{\beta}ds^{\mu\nu} \ .
\end{equation}

Therefore, the parallel transport of a vector  induces a rotation driven by curvature. For further reference let us also introduce here the homothetic curvature tensor, defined as

\begin{equation} \label{eq:hc}
g^{\alpha\beta}R_{\alpha\beta\mu\nu}=R^{\alpha}_{\;\alpha\mu\nu}\equiv\Omega_{\mu\nu} \ ,
\end{equation}
which shall be useful later. The curvature can be decomposed into eleven  irreducible components. In~the language of forms, six of these belong to the antisymmetric part of the curvature 2-form $R_{[ab]}$ (36~components) while the other five constitute the symmetric part $R_{(ab)}$ (60 components), which~vanishes in the absence of non-metricity.

\subsubsection{Torsion}

The torsion tensor can be introduced in holonomic coordinates as the antisymmetric part of the affine connection $\Gamma^{\alpha}_{\;[\beta\gamma]}$. It has 24 independent components given by

\begin{equation}
T^{\alpha}_{\;\beta\gamma}\equiv \Gamma^{\alpha}_{\;[\beta\gamma]} \ .
\end{equation}

If we apply it to the vectors $U$ and $V$ we obtain the new vector field

\begin{equation}
T(U,V)=\nabla_{U}V-\nabla_{V}U-[U,V] \ .
\end{equation}

This expression allows to visualize the geometrical effect of torsion. Indeed, starting from a given point on the manifold and parallel-transporting $V$ along the (integral curve of) $U$, through~an infinitesimal distance, and doing the complementary trip with $U$, that is, parallel-transporting it along~$V$, the expected parallelogram does not close if the manifold has a non-vanishing torsion. The~two end points are separated from each other by a (spacetime) translation, given by

\begin{equation}
\xi^{\alpha}\approx 2T^{\alpha}_{\;\mu\nu}ds^{\mu\nu} \ .
\end{equation}

The torsion field has three irreducible pieces completing the 24 independent components, $T^{\lambda}_{\;\mu\nu}=\bar{T}^{\lambda}_{\;\mu\nu}+\tfrac{2}{3}\delta^{\lambda}_{[\nu}T_{\mu]}+g^{\lambda\sigma}\epsilon_{\mu\nu\sigma\rho} \breve{T}^{\rho}$, where the traceless tensor (16 components) obeys $\bar{T}^{\lambda}_{\;\mu\lambda}=0$ and $\epsilon^{\lambda\mu\nu\rho}\bar{T}_{\mu\nu\rho}=0$, while $T_{\mu}$ is the trace vector and $\breve{T}^{\lambda}\equiv \tfrac{1}{6}\epsilon^{\lambda\alpha\beta\gamma}T_{\alpha\beta\gamma}$ the  pseudo-trace (axial) vector.

\subsubsection{Non-Metricity}

In holonomic coordinates the non-metricity tensor can be defined as the covariant derivative of the spacetime metric $g_{\beta\gamma}$. It has 40 independent components

\begin{equation}
Q_{\alpha\beta\gamma}=\nabla_{\alpha}g_{\beta\gamma} \ .
\end{equation}

The trace-vector part  $Q_{\mu}\equiv Q_{\mu\alpha}^{\;\;\;\;\alpha}$ is the only non-vanishing part of non-metricity in a Weyl geometry (e.g., Riemann-Weyl or Cartan-Weyl, to be discussed later), known as the Weyl co-vector. It~is actually related to the homothetic curvature (\ref{eq:hc}) via the equation

\begin{equation}
\Omega_{\mu\nu}=-\dfrac{1}{2}\left(\partial_{\mu}Q_{\nu}-\partial_{\nu}Q_{\mu}\right) \ .
\end{equation}

After the parallel transport of a vector along an infinitesimal closed loop, there is a change in length given by

\begin{equation}
\delta l\approx l(v)\Omega_{\mu\nu}ds^{\mu\nu} \ .
\end{equation}

Indeed, if  $Q_{\mu}$ is the only non-vanishing part of $Q_{\alpha\beta\gamma}$, then $\nabla_{\alpha}g_{\beta\gamma}\sim Q_{\alpha}g_{\beta\gamma}$ and the norm of vectors change due to this Weyl co-vector as

\begin{equation}
\nabla_{\nu}(V^{2})\sim Q_{\nu}V^{2},\qquad V^{2}=g_{\beta\gamma}V^{\beta} V^{\gamma} \ .
\end{equation}

Therefore, under the presence of non-metricity the parallel transport of a vector involves a change on its length.

From the definition of the non-metricity tensor, one can derive the following Bianchi identity-type~relation:

\begin{equation}
\nabla_{[\mu}Q_{\nu]\alpha\beta}=-R_{(\alpha\beta)\mu\nu}+Q_{\lambda\alpha\beta}T^{\lambda}_{\;\;\mu\nu} \ . \label{bianchinonmetricity}
\end{equation}

Therefore, if $Q_{\alpha\beta\gamma}\neq 0$ then $R_{(\alpha\beta)\mu\nu}\neq 0$. The quantity $R_{(\alpha\beta)\mu\nu}$ can be identified as the non-Riemannian part of the curvature, and its related to a non-Lorentzian linear connection and the breaking of Lorentz invariance. Finally, non-metricity can be decomposed into its trace-vector part and the traceless tensor part $\bar{Q}_{\alpha\beta\gamma}$, according to $Q_{\alpha\beta\gamma}=\bar{Q}_{\alpha\beta\gamma}+\tfrac{1}{4}g_{\beta\gamma}Q_{\alpha}$. In the language of forms, the~latter can be further decomposed into a shear co-vector and a shear 2-form part in such a way that at the end of the day the tensor valued non-metricity 1-form has four irreducible pieces with respect to the Lorentz group.

To summarize the geometrical interpretation of the three pieces of the connection, one can say that they are associated to changes in the properties of a vector when parallel-transported: curvature yields a rotation, torsion a non-closure of its parallelograms, and non-metricity  a change on its length. In turn, these geometrical interpretations have a deep impact in the physical models built upon any such pieces, as we learned decades ago from the physics of solid state systems with defects \cite{KittelBook}.

\subsection{A Brief Note on the Conformal and Metric Structures of Spacetime Geometry}

Important developments in geometrical methods in field theories suggest that the spacetime metric might not be considered as a fundamental field. In fact, the conformally invariant part of the metric can be derived from local and linear electrodynamics \cite{Hehl:2005hu,Hehl:1999bt,Lammerzahl:2004ww,Itin:2004qr, Hehl:2004yk}. The pre-metric approach to electrodynamics give fully general, coordinate-free, covariant inhomogeneous equations

\begin{equation}
dH=J \ ,
\end{equation}
 and homogeneous field equations

\begin{equation}
dF=0 \ ,
\end{equation}
from charge conservation and magnetic flux conservation, respectively. Since there is no metric involved, electrodynamics is not linked to a Minkowski spacetime at a fundamental level. In order to close the system the postulate on the constitutive relations $H=H(F)$ is required which, in vacuum, can be interpreted as constitutive relations for the spacetime itself \cite{Cabral:2016yxh}. On the other hand, linear, local and homogeneous constitutive relations $H=\lambda\star F$ introduce the spacetime metric\footnote{In $d$ dimensions the Hodge star operator maps $p$-forms to $(d-p)$-forms. In components we get $H_{\mu\nu}=\tfrac{\lambda}{2}\sqrt{-g}g^{\alpha\lambda}g^{\beta\gamma}\epsilon_{\mu\nu\alpha\beta}F_{\lambda\gamma}$.}, which is also involved in the more general linear-type relations $H_{\alpha\beta}=\kappa^{\mu\nu}_{\alpha\beta}F_{\mu\nu}$. Since the excitation $H$ and the field strength $F$ are 2-forms, the tensor $\kappa^{\mu\nu}_{\alpha\beta}$ has 36 independent components characterizing the electromagnetic (propagation) properties of spacetime.

From pre-metric electrodynamics plus linear, local, homogeneous constitutive relations, one can derive  the conformally invariant part of the spacetime metric. Once the propagation of electromagnetic fields is considered and the geometrical optics limit is taken, one finds a quartic Fresnell surface which becomes a quadratic surface under the imposition of zero birefringence (double refraction) in vacuum. This quadratic surface defines the light-cone. Therefore, pre-metric electrodynamics, together with linear, local, homogeneous constitutive spacetime-electromagnetic  relations and zero birefringence, gives the spacetime metric up to a conformal factor. The conformally invariant part of the metric (the~causal structure) is derived from linear electrodynamics.

We can conclude this part by stating that the resulting light cone or causal-electromagnetic structure is a conformal geometry with local conformal symmetries associated to the light cone at each spacetime point. In such a geometry, if one parallel-transports one light cone from a given point to a neighbouring point, it will be deformed according to the non-metricity tensor, which is linked to the existence of a non-Lorentzian linear connection ($\Gamma^{ab}\neq -\Gamma^{ba}$). This alone is sufficient to break Lorentz symmetry. One sees very clearly that, in addition to the metric not being fundamental, there is a primacy of the conformal (over the metric) structure, which itself is an electromagnetic derived quantity. Moreover, electrodynamics is fundamentally connected to the conformal geometrical structure (and  to the conformal group), but neither to the Poincar\'{e}/Lorentz group nor to Minkowski spacetime.

\section{Metric-affine Formalism and Classical Spacetime\label{sec:III}}
\unskip

\subsection{A Brief Outlook on Metric-Affine Gravity}

Let us now summarize the fundamental structures of spacetime and their relation to symmetry groups. We have the fundamental 1-forms, the linear co-frame $\theta^{a}$ and the linear connection $\Gamma^{a}_{\;\;b}$, which~are the potentials for the 4-dimensional translations $T(4)$ and the general linear $GL(4,\Re)$ groups, respectively. The corresponding field strengths correspond to well known objects from differential geometry, namely, the torsion $T^{a}=D\theta^{a}=d\theta^{a}+\Gamma^{a}_{\;\;b}\wedge\theta^{b}$ and curvature $R^{a}_{\;\;b}=d\Gamma^{a}_{\;\;b}+\Gamma^{a}_{\;\;c}\wedge \Gamma^{c}_{\;\;b}$ 2-forms, respectively. The metric is introduced as the 0-form potential with the corresponding field strength being the non-metricity 1-form $Q_{ab}=Dg_{ab}$. The linear frame establishes a link between the (symmetries on the) local tangent fibers and the spacetime manifold while the linear connection  can be viewed as a guidance field reflecting the inertial character for matter fields propagating on the spacetime manifold. Finally, the metric allows the determination of spatial and temporal distances and angles. The spacetime geometry with all these structures is called a \emph{metric-affine geometry} (MAG), with~non-vanishing torsion, curvature and non-metricity, and its fundamental local group of spacetime symmetries is the affine group
 $A(4,\Re)=T(4)\rtimes GL(4,\Re)$, which is the semi-direct product of the group of translations and the general linear group.

A truly independent linear connection is given by the decomposition in (\ref{independentconnection}) that is useful to analyse the relation between non-Lorentzian metrics, the breaking of Lorentz invariance and the presence of non-metricity. Since $Q_{ab}=2N_{(ab)}$, if non-metricity is zero, then the connection is Lorentzian and the spacetime geometry is the RC one with curvature and torsion. Such a spacetime is fundamentally linked to the local symmetry group of Poincar\'{e} transformations $P(1,3)=T(4)\rtimes SO(1,3)$, which is the semi-direct product between the translations $T(4)$ and the Lorentz group $SO(1,3)$. As one can see from the expressions for the field strengths (torsion and curvature), the connection also enters in the expression for the field strength of the co-frame. This term is unavoidably present and is due to the semi-direct product\footnote{The semi direct product implies that the generators of $T(4)$ and $GL(4,\Re)$ (or $SO(1,3)$) do not commute.} structure of the Poincar\'{e} group (or the affine group) and, therefore, curvature and torsion are somehow intertwined.

In the self-consistent metric-affine formalism, the gravitational interaction is described as a gauge theory of the affine group $A(4,\Re)$, together with the assumption of a metric, with the  potentials ($\theta^{a},\Gamma^{ab}$) coupled to the corresponding Noether currents ($\tau^{a},\Delta^{a}_{\;\;b}$). The latter are the vector-valued canonical energy-momentum $\tau^{a}=\delta \mathcal{L}_{mat}/\delta\theta_{a}$ and the tensor-valued hypermomentum $\Delta^{a}_{\;b}=\delta \mathcal{L}_{mat}/\delta\Gamma_{a}^{\;b}$ 3-form currents\footnote{These currents can be represented as 1-forms or as 3-forms. In fact, in the gauge approach to gravity they emerge naturally from Noether equations as 3-forms, being natural objects for integration over volumes. As 1-forms one can write $\tau^{a}=\tau^{a}_{\;\mu}dx^{\mu}$ and $\Delta^{a}_{\;\;b}=\Delta^{a}_{\;\;b\mu}dx^{\mu}$. This map between 3-form or 1-form representation is related to the fact that in $d$-dimension a $k$-form has $\tfrac{d!}{k!(d-k)!}$ independent components. In four dimensions both 3-forms and 1-forms have four independent components.}, respectively, while the metric $g_{ab}$ couples to the symmetric (Hilbert) energy momentum $T_{ab}=2\delta \mathcal{L}_{mat}/\delta g^{ab}$. The hypermomentum can be decomposed according to the expression

\begin{equation} \label{eq:hypermomentum}
\Delta_{ab}=s_{ab}+\dfrac{1}{4}g_{ab}\Delta^{c}_{\;\;c}+\bar{\Delta}_{ab} \ ,
\end{equation}
including the spin $s_{ab}=-s_{ba}$, the dilatation $\Delta^{c}_{\;\;c}$, and the shear $\bar{\Delta}_{ab}$ currents.
The Noether currents are the fundamental sources of MAG. 
A truly independent connection in MAG can be written as
\begin{equation}
\Gamma^{ab}=\Gamma^{[ab]}+\dfrac{1}{4}g^{ab}\Gamma^{c}{}_{c}+\left(\Gamma^{ab}-\dfrac{1}{4}g^{ab}\Gamma^{c}{}_{c}\right),
\label{MAGconnection}
\end{equation}
including the Lorentzian piece $\Gamma^{[ab]}$, the trace part $\dfrac{1}{4}g^{ab}\Gamma^{c}{}_{c}$ and the shear part $\Gamma^{ab}-\dfrac{1}{4}g^{ab}\Gamma^{c}{}_{c}$. The~Lorentzian connection couples to the spin current and the trace and shear parts couple to the dilatation and shear currents, respectively. As previously said it is the non-Lorentzian part of the connection that imply the non-vanishing of non-metricity, therefore, the dilatation and shear  hypermomentum currents are intimately related to the non-metricity of metric affine spacetime~geometry.

The variational principle is applied to the action of this theory (including the gravitational part and the matter Lagrangian) by varying it with respect to the gauge potentials of the affine group, ($\theta^{a},\Gamma^{ab}$). This leads to two sets of dynamical equations, while a third set of equations is obtained by varying the action with respect to the metric potential $g_{ab}$. At the end, the dynamics is described only via two sets of equations, since the gravitational equation obtained by varying with respect to the metric potential or the one obtained by variation with respect to the translational potential can be dropped out, as long as the other gravitational equation (derived from variation with respect to the linear connection) is fulfilled. This procedure and the fundamental quantities and relations here exposed summarizes the basics of the MAG formalism.

\subsection{Classical Spacetime Paradigms}

We now revise some important classical spacetime paradigms, which play a fundamental role in the formulation of gravitational theories.

\subsubsection{Minkowski Spacetime---$M_{4}$}

This is a pseudo-Euclidean geometry with vanishing curvature, torsion and non-metricity. It~has global/rigid Poincar\'{e} symmetries and a globally absolute causal structure.
For inertial reference systems the connection vanishes.
The inertial frame and the inertial properties of matter are defined with respect to this absolute spacetime. In particular, the mass and spin of particles can be considered to be intrinsic to particles and are classified with the help of the Casimir operators using the irreducible representations of the Poincar\'{e} group.

\subsubsection{(Pseudo)Riemann Geometry of GR---$V_{4}$}

A post-Euclidean geometry with curvature and vanishing torsion and non-metricity. It obeys local Lorentz symmetries and in local geodesic frames the (Levi-Civita) connection vanishes but (the Weyl part of) curvature is non-zero. The causal structure $ds^{2}=0$ is locally invariant under local Lorentz symmetries. The inertial properties of matter are locally defined with respect to absolute spacetime.

\subsubsection{Riemann-Cartan Geometry---$U_{4}$}

This geometry has curvature and torsion but zero non-metricity, therefore the connection is Lorentzian (spin connection). It obeys local Poincar\'{e} $P(1,3)$ symmetries and the causal structure is locally invariant under such a group. The inertial properties of matter are locally defined with respect to absolute spacetime.  The corresponding curvature and its contractions (Ricci tensor and Ricci scalar in the RC geometry) are given by the expressions
\begin{eqnarray}
R^{\alpha}_{\;\beta\mu\nu}&=&\tilde{R}^{\alpha}_{\;\beta\mu\nu}+\tilde{\nabla}_{\mu}K^{\alpha}_{\;\beta\nu}-\tilde{\nabla}_{\nu}K^{\alpha}_{\;\beta\mu}
+K^{\alpha}_{\;\lambda\mu}K^{\lambda}_{\;\beta\nu}-K^{\alpha}_{\;\lambda\nu}K^{\lambda}_{\;\beta\mu} \nonumber \\
R_{\beta\nu}&=&\tilde{R}_{\beta\nu}+\tilde{\nabla}_{\alpha}K^{\alpha}_{\;\beta\nu}-\tilde{\nabla}_{\nu}K^{\alpha}_{\;\beta\alpha} 
+K^{\alpha}_{\;\lambda\alpha}K^{\lambda}_{\;\beta\nu}-K^{\alpha}_{\;\lambda\nu}K^{\lambda}_{\;\beta\alpha}, \\
R&=&\tilde{R}-2\tilde{\nabla}^{\lambda}K^{\alpha}_{\;\lambda\alpha}+g^{\beta\nu}(K^{\alpha}_{\;\lambda\alpha}K^{\lambda}_{\;\beta\nu}-K^{\alpha}_{\;\lambda\nu}K^{\lambda}_{\;\beta\alpha}) \ , \nonumber\label{ricciRC}
\end{eqnarray}
respectively. The curvature tensor obeys the following first and second Bianchi identities
\begin{eqnarray}
\nabla_{[\gamma}R^{\alpha}_{\;\beta\vert\mu\nu]}&=&2R^{\alpha}_{\;\beta\lambda[\mu}T^{\lambda}_{\;\nu\gamma]} \,, \\
 R^{\alpha}_{\;[\beta\mu\nu]}&=&-2\nabla_{[\nu}T^{\alpha}_{\;\beta\mu]}+4T^{\alpha}_{\;\lambda[\beta}T^{\lambda}_{\;\mu\nu]} \,.
\label{BianciRC}
\end{eqnarray}

These relations can be deduced from the corresponding expressions in terms of the curvature and torsion 2-forms, namely\footnote{Here $DR^{a}_{\;\;b}=dR^{a}_{\;\;b}+\Gamma^{a}_{\;\;c}\wedge R^{c}_{\;\;b}+\Gamma^{c}_{\;\;b}\wedge R^{a}_{\;\;c}$ and $DT^{c}=dT^{c}+\Gamma^{c}_{\;\;d}\wedge T^{d}$.}
\begin{equation}
DR^{a}_{\;\;b}=0, \qquad DT^{c}= R^{c}_{\;\;d}\wedge \theta^{d}.
\end{equation}

The anholonomic or frame indices are also called symmetry indices since they are related to the tangent fibers where the local spacetime symmetries are characterized. In this case these are known also as  Lorentz indices. Spin connections are related to rotations (two Lorentz indices) and the tetrads or co-frames are related to translations (one Lorentz index). The same is valid for the corresponding field strengths. The symmetry indices have a direct link to geometrical interpretations via the strong relation between group theory and geometry. Two symmetry indices for curvature means that it is related to rotations, and one symmetry index for torsion means that it is related to translations. On~the other hand, since both curvature and torsion are represented by 2-forms, as geometrical objects these are therefore connected to 2-surfaces. One can thus say that Cartan pictured a RC geometry by associating to each infinitesimal surface element a rotation and a translation.

The components of the curvature and torsion 2-forms in terms of the tetrads and spin connection~are
\begin{eqnarray}
R^{a}_{\;b\mu\nu}&=&\partial_{\mu}\Gamma^{a}_{\;b\nu}-\partial_{\nu}\Gamma^{a}_{\;b\mu}+\Gamma^{a}_{\;c\mu}\Gamma^{c}_{\;b\nu}-\Gamma^{a}_{\;d\nu}\Gamma^{d}_{\;b\mu}\ ,
\label{curvconexaospin} \\
T^{a}_{\;\mu\nu}&=&\partial_{\mu}\theta^{a}_{\;\nu}-\partial_{\mu}\theta^{a}_{\;\nu}+\Gamma^{a}_{\;b\mu}\theta^{b}_{\;\nu}-\Gamma^{a}_{\;b\nu}\theta^{b}_{\;\mu} \ , \label{torsaotetradas}
\end{eqnarray}
respectively. On the other hand, the components of the 1-form spin connection

\begin{equation}
\Gamma^{a}_{\;b\nu}=\tilde{\Gamma}^{a}_{\;b\nu}+K^{a}_{\;b\nu} \ ,\label{conexaospinRC}
\end{equation}
where $K^{a}_{\;b\mu}$ are the components of the contortion 1-form, are related to the  holonomic (spacetime) components of the affine connection through the relations
\begin{eqnarray}
\Gamma^{a}_{\;b\nu}&=& \theta^{a}_{\;\mu}\partial_{\nu}e_{b}^{\;\mu}+\theta^{a}_{\;\mu}\Gamma^{\mu}_{\;\beta\nu}e_{b}^{\;\beta} \\ \Gamma^{\lambda}_{\;\nu\mu}&=&e_{a}^{\;\lambda}\partial_{\mu}\theta^{a}_{\;\nu}+e_{a}^{\;\lambda}\Gamma^{a}_{\;b\mu} \theta^{b}_{\;\nu} \ .\label{conexaospinafim}
\end{eqnarray}

Since the connection characterizes the way the linear frame/co-frame changes from point to point, these relations can be deduced from the equation

\begin{equation}
 \partial_{\nu}e_{b}^{\;\mu}+\Gamma^{\mu}_{\;\beta\nu}e_{b}^{\;\beta}\equiv \Gamma^{c}_{\;b\nu}e_{c}^{\;\mu} \ ,
\end{equation}
 which expresses the fact that the total covariant derivative of the tetrads with respect to both holonomic and anholonomic (Lorentz) indices is vanishing, that is, $\partial_{\nu}e_{b}^{\;\mu}+\Gamma^{\mu}_{\;\beta\nu}e_{b}^{\;\beta}- \Gamma^{c}_{\;b\nu}e_{c}^{\;\mu}=0$ and $\partial_{\mu}\theta^{a}_{\;\nu}-\Gamma^{\beta}_{\;\nu\mu}\theta^{a}_{\;\beta}+ \Gamma^{a}_{\;b\mu}\theta^{b}_{\;\nu}=0$. By changing from one spacetime point to another in the frame/coframe, the~tetrads also change. The new tetrads $\bar{e}_{b}^{\;\mu}$ can be expressed in terms of the original ones as $\bar{e}_{b}^{\;\mu}=\Lambda^{a}_{b}e_{a}^{\;\mu}$, and by using the relations

\begin{equation}
\eta_{ab}=e_{a}^{\;\mu}e_{b}^{\;\nu}g_{\mu\nu}, \quad g_{\mu\nu}=\theta^{a}_{\;\mu}\theta^{b}_{\;\nu}\eta_{ab}, \quad \sqrt{-g}=det(\theta^{a}_{\;\mu}), \label{tetradasmetrica}
\end{equation}
one can show that  $\Lambda^{a}_{b}$ are Lorentz matrices, that is, the tetrads undergoes a Lorentz rotation under the motion from one spacetime point to another. For that reason, the linear connection, which characterizes the change in the frame/co-frame, is called Lorentzian or spin connection. As previously said, the six Lorentzian connections are the potentials associated to the generators of the Lorentz group.

The matrices $e_{a}^{\;\mu}$ and their inverses, which establish a correspondence between the local spacetime metric and the Minkowski metric on the tangent/cotangent planes, constitute a map between holonomic and anholonomic bases. For example, for vectors

\begin{equation}
v^{\mu}=v^{a}e_{a}^{\;\mu} , \quad v^{b}=v^{\nu}\theta^{b}_{\;\nu} , \quad  h_{\mu}=h_{a}\theta^{a}_{\;\mu} , \quad h_{b}=h_{\nu}e_{b}^{\;\nu}.
\end{equation}

Since the tangent and co-tangent spaces, $T_{p}(\mathcal{M})$ and  $T^{*}_{p}(\mathcal{M})$, change while moving from one point on the manifold to another, the notion of covariant derivative is extended for quantities with Lorentz indices. This is done with the spin connection. For (Lorentz) vectors and co-vectors we have

\begin{equation}
^{s}\nabla_{\mu}v^{a}=\partial_{\mu}v^{a}+\Gamma^{a}_{\;b\mu}v^{b}, \quad  ^{s}\nabla_{\nu}h_{c}=\partial_{\nu}h_{c}-\Gamma^{d}_{\;c\nu}h_{d}.
\end{equation}

The anholonomic basis $ {e_{a}}$ and ${\theta^{b}}$ also characterize the local spacetime symmetries in the tangent fibers, since the local symmetry group algebra is implicit in the relations

\begin{equation}
[ e_{a}, e_{b}]=f^{c}_{\;ab} e_{c},\qquad f^{c}_{\;ab}=e_{a}^{\;\mu}e_{b}^{\;\nu}(\partial_{\nu}\theta^{c}_{\;\mu}-\partial_{\mu}\theta^{c}_{\;\nu}),\label{structconst}
\end{equation}
where the  (group) structure constants $f^{a}_{\;bc}$ are known as Ricci rotation coefficients.

In RC geometry the affine (self-parallel) geodesics differ, in general, from the extremal geodesics and are given by

\begin{equation}
\frac{du^{\alpha}}{ds}+\Gamma^{\alpha}_{\;\beta\gamma}u^{\beta}u^{\gamma}=0 \quad \Leftrightarrow \quad \frac{du^{\alpha}}{ds}+\tilde{\Gamma}^{\alpha}_{\;\beta\gamma}u^{\beta}u^{\gamma}=-K^{\alpha}_{\;(\beta\gamma)}u^{\beta}u^{\gamma},
\end{equation}
respectively, where $u^{\alpha}$ are the components of the (4-velocity) tangent vector to the curve. However, in~the RC spacetime matter particles with vanishing intrinsic spin are insensitive to the non-Riemannian part of the geometry and, therefore, to torsion. Such particles follow the extremal paths computed from the Levi-Civita connection. Moreover if torsion is completely antisymmetric, as in the case of ECSK theory, then $K^{\alpha}_{\;(\beta\gamma)}=0$ and the extremal and self-parallel geodesics coincide. In any case, the~appropriate evaluation of the motion of particles with intrinsic spin (fermions) in a RC spacetime should be performed from an analysis of the corresponding Dirac equation and then proceeding with a classical approximation, for instance, using a WKB method.

As we mentioned before, since curvature and torsion are 2-forms, one can imagine the RC geometry as having at each point an associated infinitesimal surface element with a rotational (curvature) and translational (torsion) transformation.~The picture of a RC manifold with a discrete structure can naturally emerge from imposing a finite minimum surface element (rater than infinitesimal elements), which would imply, to some extent, that torsion and curvature would become quantized. Indeed, as we shall see, the mathematical methods of Cartan's exterior calculus  (differential forms) within gauge theories of gravity contribute to clarify the appropriate physical degrees of freedom and mathematical objects that should be quantized in a quantum (Yang-Mills) gauge theory of gravity. This procedure, in this formalism of forms, can be done in a metric-free (pre-metric) way, avoiding the difficulties often found in perturbative approaches that require some (well behaved) background (spacetime-vacuum), with respect to which the perturbations can be defined. Since any such gravity theory should be (metric) background independent, this is usually a huge problem, since the background is also the very thing one would like to quantize and should come as a solution of the dynamical equations. Gravity in exterior forms can shed some light on this challenge, at least by establishing a metric-independent framework and a well identified set of canonically conjugate variables to be quantized. As we shall see later, in Yang-Mills type of gauge theories of gravity this will not require the full metric structure, but only the conformally invariant part of the metric, that is the conformal-causal structure. The latter is introduced via the Hodge star operator in the constitutive relations between the field strengths and the conjugate momenta.

\subsubsection{Riemann-Weyl Geometry---$W_{4}$}

This is the spacetime geometry implicit in Weyl's gauge theory for unifying gravity and electromagnetism by extending the local Lorentz symmetries to include dilatations. It has curvature and (the trace-vector part of) non-metricity but vanishing torsion, and it obeys local symmetries under the Weyl group $W(1,3)$, which includes the $P(1,3)$ group and dilatations. The causal structure is locally invariant under the $W(1,3)$ group, but the spacetime (metric) is not absolute, changing under dilatation-type of coordinate transformations.
Accordingly, the inertial properties of matter cannot be defined with respect to an absolute metric structure. One may postulate that matter is endowed with conformally-invariant physical properties.

The Weyl connection is given by $
\Gamma^{\alpha}_{\;\beta\nu}=\tilde{\Gamma}^{\alpha}_{\;\beta\nu}+q^{\alpha}_{\;\beta\nu}$, where the distortion tensor in this case is related to the Weyl co-vector $Q_{\mu}\equiv Q_{\mu\lambda}^{\quad \lambda}$ as

\begin{equation}
q^{\alpha}_{\;\beta\nu}\equiv\frac{1}{2}(\delta^{\alpha}_{\beta}Q_{\gamma}+\delta^{\alpha}_{\gamma}Q_{\beta}-Q^{\alpha}g_{\beta\gamma}) \ ,
\end{equation}
while the curvature of the Weyl connection is given by

\begin{equation}
R^{\alpha}_{\;\beta\mu\nu}=\tilde{R}^{\alpha}_{\;\beta\mu\nu}+\tilde{\nabla}_{\mu}q^{\alpha}_{\;\beta\nu}-\tilde{\nabla}_{\nu}q^{\alpha}_{\;\beta\mu}+q^{\alpha}_{\;\lambda\mu}q^{\lambda}_{\;\beta\nu}-q^{\alpha}_{\;\lambda\nu}q^{\lambda}_{\;\beta\mu}.
\end{equation}

The Weyl co-vector is the trace vector part of the non-metricity and, therefore, there are scale (dilatations) type of distortions in the geometry, implicit in the relation $\nabla_{\alpha}g_{\beta\gamma}\sim Q_{\alpha}g_{\beta\gamma}$ and, as~previously mentioned, the length of vectors change in this geometry as $\nabla_{\nu}(V^{2})\sim Q_{\nu}V^{2}$.
Weyl~identified the Weyl co-vector as the electromagnetic 4-potential, and the previously defined homothetic curvature (\ref{eq:hc}) as the electromagnetic Faraday tensor. We stress that the pre-metric foundations of electrodynamics indeed show the  deep connection between electromagnetism and conformal geometry and  conformal symmetries, which presuppose a natural framework for the breaking of Lorentz symmetry.

\subsubsection{Riemann-Cartan-Weyl Geometry---$Y_{4}$}

This is a generalization of the Weyl geometry by including torsion, which is therefore the appropriate manifold for the Weyl gauge theories of gravity. Only matter with hypermomentum~(\ref{eq:hypermomentum}), that includes the spin current and the dilatations current, can be sensitive to torsion and to the (Weyl-covector part of) non-metricity.

\subsubsection{Metric-affine Geometry ($L_{4},g$)}

This is the richest kind of geometry considered here, having non-vanishing curvature, torsion and non-metricity. The natural symmetry group is the affine group $A(4,\Re)$. The affine connection can be written as $\Gamma^{\alpha}_{\;\beta\nu}=\tilde{\Gamma}^{\alpha}_{\;\beta\nu}+N^{\alpha}_{\;\beta\nu}$, where the distortion tensor $N^{\alpha}_{\;\beta\nu}=K^{\alpha}_{\;\beta\nu}+L^{\alpha}_{\;\beta\nu}$ includes both contortion and  disformation, as described in Section \ref{eq:section2b}. The curvature can be written as

\begin{equation}
R^{\alpha}_{\;\beta\mu\nu}=\tilde{R}^{\alpha}_{\;\beta\mu\nu}+\tilde{\nabla}_{\mu}N^{\alpha}_{\;\beta\nu}-\tilde{\nabla}_{\nu}N^{\alpha}_{\;\beta\mu}+N^{\alpha}_{\;\lambda\mu}N^{\lambda}_{\;\beta\nu}-N^{\alpha}_{\;\lambda\nu}N^{\lambda}_{\;\beta\mu},
\end{equation}
or, alternatively, as

\begin{equation}
R^{\alpha}_{\;\beta\mu\nu}=\bar{R}^{\alpha}_{\;\beta\mu\nu}+\bar{\nabla}_{\mu}L^{\alpha}_{\;\beta\nu}-\bar{\nabla}_{\nu}L^{\alpha}_{\;\beta\mu}+L^{\alpha}_{\;\lambda\mu}L^{\lambda}_{\;\beta\nu}-L^{\alpha}_{\;\lambda\nu}L^{\lambda}_{\;\beta\mu},
\end{equation}
where $\bar{R}^{\alpha}_{\;\beta\mu\nu}$ and $\bar{\nabla}_{\mu}$ are the curvature and covariant derivative of a Riemann-Cartan connection.

The above discussion represents just a quick excursion in the \emph{affinesia} world\footnote{This anecdotal notion was first coined by Jos\'e Beltr\'an-Jim\'enez in one of his talks about the trinity of gravity \cite{BeltranJimenez:2019tjy}.}, where different geometries emerge depending on the pieces of the connection one decides to keep/remove. In the next section, we shall draw our attention to the gauge formulation of gravity and its relation to all these spacetime geometries.

\section{Gauge Theories of Gravity}\label{sec:IV}

The gauge approach to gravity broadens our study of the deep relation between symmetry principles (group theory) and geometrical methods. Of special relevance for our analysis is the PGTG class, which constitutes a promising candidate for an appropriate description of classical gravity including post-Einstein strong-field predictions. In order to present the structure of the gauge approach to gravity, it is useful first to revisit the Weyl-Yang-Mills  formalism for gauge fields.

\subsection{The Weyl-Yang-Mills Formalism}

The gauge approach to Yang-Mills fields follows from two major steps, the rigid (global) symmetries of a physical system described by a matter Lagrangian, and the localization (gauging) of those symmetries.

In the first step one considers rigid (global) symmetries as follows:
\begin{itemize}[align=parleft,leftmargin=*,labelsep=4.99mm]
\item Start with a field theory: $\mathcal{L}_{m}=\mathcal{L}_{m}(\Psi,d\Psi)$ of some matter fields $\Psi$.
\item The matter Lagrangian is invariant under some internal symmetry group, described by a (semi-simple) Lie group with generators $T_{a}$.
\item  Noether's first theorem implies a conserved current: $dJ=0$.
\end{itemize}

In the second step the localization (gauging) of the symmetries is performed according to the following procedure:
\begin{itemize}[align=parleft,leftmargin=*,labelsep=4.99mm]
\item The symmetries are described on each spacetime point introducing the compensating (gauge) field $A=A^{a}_{\mu}T_{a}dx^{\mu}$.
\item This is a new field that couples minimally to matter and represents a new interaction.
\item To preserve the symmetries this gauge potential $A$ transforms in a suitable way allowing to construct a (gauge) covariant derivation $d\Psi\longrightarrow D\Psi=(d+A)\Psi$.
\item The Lagrangian includes this minimal coupling between the matter fields and the gauge potential $\mathcal{L}(\Psi,d\Psi)\longrightarrow \mathcal{L}(\Psi,D\Psi)$.
\item The gauge potential acts on the components of the matter fields defined with respect to some reference frame. Geometrically, it is the connection of the frame bundle (fiber bundle) related to the symmetry group.
\item The conservation equation is generalized as $dJ=0\rightarrow DJ=0$.
\end{itemize}

In order for $A$ to represent a true dynamical variable with its own degrees of freedom, the~Lagrangian of the theory has to include a kinetic term, representing the new interaction $\mathcal{L}=\mathcal{L}_{m}+\mathcal{L}_{A}$.  The invariance of $\mathcal{L}_A$ is secured by constructing it with the gauge invariant field strength $F=DA=dA+A\wedge A$ which, geometrically, can be interpreted as the curvature 2-form of the fiber bundle. Note that to have second order (on $A$) inhomogeneous Yang-Mills field equations, one must choose $\mathcal{L}_{A}=\mathcal{L}_{A}(F)$.

Written in terms of exterior forms, the inhomogeneous Yang-Mills equations for the gauge potential are

\begin{equation}
DH=J \ ,
\end{equation}
where $H=\partial \mathcal{L}/\partial F$ is the excitation 2-form, and $J=\partial \mathcal{L}_{m}/\partial A$ is the conserved Noether current, which acts as a source for the potential. Here, $DH\equiv dH+A\wedge H$. The homogeneous field equation corresponds to a Bianchi identity, obtained from the derivation of the potential twice, namely

\begin{equation}
DF=0 \quad  \Leftrightarrow \quad dF=-A\wedge F \ .
\end{equation}

Note also that the equation for the conservation of the Noether current is generalized via the gauge covariant exterior derivative

\begin{equation}
DJ=0 \quad  \Leftrightarrow \quad dJ=-A\wedge J \ .
\end{equation}

For non-abelian  groups the gauge field contributes with an associated (``isospin") current, $-A\wedge H$. In such a case  $dJ\neq 0$, and the (gauge) interaction field is charged, unlike the case of abelian groups, such as the $U(1)$ group of electromagnetism.

In order to have a wave-like inhomogeneous Yang-Mills (quasi-linear) equation, and paralleling the case of electromagnetism, $\mathcal{L}$ can depend  quadratically with $F$ at most and, therefore, $H$ must depend linearly, for instance as $H=H(F)=\alpha\star F$. The Yang-Mills inhomogeneous equations then turn into

\begin{equation}
D\star F=d\star F+A\wedge\star F=\alpha^{-1}J \Leftrightarrow d\star F=\alpha^{-1}(J+ J^{A}),
\end{equation}
where $J^{A}\equiv -\alpha  A\wedge\star F$.
In this formalism one can see the clear analogies between classical mechanics and Yang-Mills field theory, which we summarize in Table \ref{table:I}. In particular, it is clear that the field strength $F$ is the generalized velocity while the excitation $H$ is the conjugate momentum. The~constitutive relation $H(F)$  is implicit in the Lagrangian formulation and corresponds in perfect analogy to the functional relation between generalized velocities and conjugate momenta of classical mechanics. As an extension of these analogies, one is naturally led to identify the appropriate pairs of canonically conjugate variables that should be quantized in the corresponding (canonical) quantum
field theory (see Table \ref{table:II}).

\begin{table}[H]
\centering
\caption{The analogies between classical mechanics, Yang-Mills field and gravity-Yang-Mills theory in the language of exterior forms.}
\scalebox{.9}[0.9]{\begin{tabular}{cccc}
\toprule
        &\textbf{Gravity Yang-Mills} & \textbf{Yang-Mills} & \textbf{Classical Mechanics}  \\
        && \boldmath{$\mathcal{L}=\mathcal{L}(A,DA)$} & \boldmath{$\mathcal{L}=\mathcal{L}(q,\dot{q})$}  \\
\midrule

Configuration &$(\Gamma^{a}_{b},\vartheta^{a})$& $A$ & $q$ \\
  variables &&  &   \\
\midrule

\multirow{2}{*}{Generalized velocities}&$R^{a}_{b}=D\Gamma^{a}_{b}$& $F\equiv DA$& $\dot{q}$ \\
       &$T^{a}=D\vartheta^{a}$&  &  \\
\midrule
\multirow{2}{*}{Lagrange  equations} &$D\left(\frac{\partial \mathcal{L}}{\partial R^{ab}} \right)-\varsigma_{ab}=s_{ab}$& $D\left(\frac{\partial \mathcal{L}}{\partial F} \right) =J$& $\frac{d}{dt}\left(\frac{\partial \mathcal{L}}{\partial \dot{q}} \right)=\frac{\partial \mathcal{L}}{\partial q}$ \\
     &$D\left(\frac{\partial \mathcal{L}}{\partial T^{a}} \right)-\pi_{a}=\tau_{a}$&$J=\frac{\partial \mathcal{L}_{m}}{\partial A}$  &  \\
\midrule
\multirow{2}{*}{Conjugate momenta}  &$H_{ab}=\left(\frac{\partial \mathcal{L}}{\partial R^{ab}} \right)$& $H=\frac{\partial \mathcal{L}}{\partial DA}=\frac{\partial \mathcal{L}}{\partial F}$& $p=\frac{\partial \mathcal{L}}{\partial \dot{q}}$ \\
     &$H_{a}=\left(\frac{\partial \mathcal{L}}{\partial T^{a}} \right)$& &  \\
\midrule

\multirow{2}{*}{Constitutive  relations}&$H^{ab}=H^{ab}(R^{ab},T^{a})\qquad H^{a}=H^{a}(R^{ab},T^{a})$& $H=H(F)$& $p=p(\dot{q})$ \\
       &$H^{ab}\sim \star R^{ab}\qquad H^{a}\sim \star T^{a}$ (linear)&$H\sim\star F$ (linear) &   \\
\midrule

\multirow{2}{*}{Canonical   variables}  &$(\Gamma^{a}_{b},H^{a}_{b})\qquad(\vartheta^{a},H^{a})$& $(A,H)$& $(q,p)$ \\
&& &   \\
\midrule

\multirow{2}{*}{Hamiltonian } &$\mathcal{H}\equiv R^{ab}\wedge H_{ab}+T^{a}\wedge H_{a}-\mathcal{L}(\Gamma,\vartheta,R,T)$& $\mathcal{H}\equiv F\wedge H-\mathcal{L}$& $\mathcal{H}\equiv \dot{q}p-\mathcal{L}(q,\dot{q})$ \\
     &$R^{ab}=R^{ab}(H^{ab},H^{a})\qquad T^{a}=T^{a}(H^{ab},H^{a})$&$F=F(H)$  & $\dot{q}=\dot{q}(p)$ \\
\midrule
\multirow{2}{*}{Hamilton  equations}  &$D\Gamma^{ab}=\frac{\partial\mathcal{H}}{\partial H_{ab}} \qquad DH_{ab}=-\frac{\partial \mathcal{H}^{eff}}{\partial \Gamma^{ab}}$& $DA=F=\frac{\partial \mathcal{H}}{\partial H}$& $\frac{d}{dt}q=\frac{\partial \mathcal{H} }{\partial p}\qquad\frac{d}{dt}p=-\frac{\partial \mathcal{H}}{\partial q}$ \\
     &$D\vartheta^{a}=\frac{\partial\mathcal{H}}{\partial H_{a}} \qquad DH_{a}=-\frac{\partial \mathcal{H}^{eff}}{\partial \vartheta^{a}}$&$DH=-\frac{\partial \mathcal{H}^{m}}{\partial A}=J$ &   \\
\bottomrule

\end{tabular}}
\label{table:I}
\end{table}
\unskip

\begin{table}[H]
\centering
\caption{The analogies between canonical quantization in quantum mechanics and in the exterior calculus approach to Yang-Mills theories and gauge theories of gravity ({\it $\grave{a}$ la} Yang-Mills).}
\begin{tabular}{cccc}
\toprule
       &\textbf{Gravity Yang-Mills} & \textbf{Yang-Mills} & \textbf{Quantum Mechanics}  \\
       &\boldmath{$\hat{\mathcal{H}}=\hat{\mathcal{H}}(\hat{\Gamma}^{ab},\hat{\vartheta}^{a},\hat{H}^{ab},\hat{H}^{a})$} & \boldmath{$\hat{\mathcal{H}}=\hat{\mathcal{H}}(\hat{A},\hat{H})$} & \boldmath{$\hat{\mathcal{H}}=\hat{\mathcal{H}}(\hat{q},\hat{p})$}  \\
\midrule

\multirow{2}{*}{Quantum operators} &$\Gamma^{ab}\rightarrow\hat{\Gamma}^{ab} \qquad \vartheta^{a}\rightarrow\hat{\vartheta}^{a}$ & $A\rightarrow \hat{A}$ & $q\rightarrow \hat{q}$ \\
 &$H^{ab}\rightarrow\hat{H}^{ab} \qquad H^{a}\rightarrow\hat{H}^{a}$  & $H\rightarrow \hat{H}\sim -i\frac{\partial}{\partial A}$ & $p\rightarrow \hat{p}= -i\hslash\frac{d}{dq}$  \\
\midrule

\multirow{2}{*}{Commutation  relations}  &$[\hat{\Gamma}^{ab},\hat{H}^{ab}]\neq 0\qquad [\hat{\vartheta}^{a},\hat{H}^{a}]\neq 0$ & $[\hat{A},\hat{H}]\neq0$& $[\hat{q},\hat{p}]=-i\hslash$ \\
    & &   &  \\
\bottomrule

\end{tabular}
\label{table:II}
\end{table}

\subsection{The Gauge Approach to Gravity}

The question that arises now is whether we can apply the same procedure to gravity. The approach of Yang, Mills and Utiyama went beyond the first ideas on gauge invariance introduced by Weyl.
In~fact, while Yang and Mills \cite{Yang:1954ek} extended Weyl's gauge principle to the $SU(2)$ isospin rotations in an attempt to describe nuclear interactions, Utiyama \cite{Utiyama:1956sy} extended the gauge principle to all semi-simple Lie groups including the Lorentz group and tried to derive GR from the gauging of the Lorentz group. Although there is some validity in his approach and an undoubtedly importance of the Lorentz group in GR, his derivation is not fully self-consistent to the formal structure of a gauge field theory. This is mainly because the Noether conserved current of the Lorentz group is not the energy-momentum, which Utiyama forced to be the source of gravity in order to obtain GR.\footnote{From a gauge theoretical perspective, the Lorentz group is not the symmetry group of Einstein's gravity. It became clear some years later that  the apropriate way to derive GR from a gauge principle is to consider it as a translational gauge theory of gravity. This class of theories lives on a Weitzenb\"{o}ck spacetime geometry with torsion and vanishing curvature and non-metricity. These geometries and its use in attempts for a unified classical field theory were worked out by Weitzenb\"{o}ck, Cartan and Einstein, for example, during the first period of the so-called teleparallel formulation gravity (up to 1938). A~second period in the 60's by Moller and others rekindled the interest in such theories which have more recently re-gained much attention, particularly via its $f(T)$ extensions (see e.g., Reference \cite{Bamba:2013jqa}).} Although these efforts did not include gravity consistently, it revealed that, on a fundamental level, gauge symmetries lie at the heart of modern field theories of physical interactions. Nevertheless, this gauge formalism eventually returned to gravity when in the 60's Sciama and Kibble localized the Poincar\'{e} group of spacetime symmetries and in this way managed to show that gravity can also be consistently described as a gauge theory. Indeed, the analogies with gauge Yang-Mills theories can be easily established, as summarized in Table \ref{table:III}.

\begin{table}[H]
\caption{The analogies between Yang-Mills fields and gravity-Yang-Mills theory in the language of exterior forms.}
\centering
\begin{tabular}{ccc}
\toprule
        &\textbf{Gravity Yang-Mills} & \textbf{Yang-Mills}   \\
        &\boldmath{$\mathcal{L}=\mathcal{L}_{g}+\mathcal{L}_{m}$}&\boldmath{$\mathcal{L}=\mathcal{L}_{A}+\mathcal{L}_{m}$}\\
        &\boldmath{$\mathcal{L}_{g}=\mathcal{L}_{g}(g_{ab},\vartheta^{a},D\Gamma^{a}_{b},D\vartheta^{a})$}& \boldmath{$\mathcal{L}=\mathcal{L}(A,DA)$}  \\
\midrule

gauge &$(\Gamma^{a}_{b},\vartheta^{a})$& $A$ \\
  potentials &1-forms& 1-form   \\
\midrule

Field &$R^{a}_{b}=D\Gamma^{a}_{b}$& $F\equiv DA$ \\
strengths &$T^{a}=D\vartheta^{a}$ &    \\
\midrule
  &$PGTG: SO(1,3)\rtimes$&  \\
Symmetry & $\qquad T(4)$ & $SU(N)$   \\
group &$MAG: GL(4,\Re)\rtimes$&  \\
 & $\qquad T(4)$ &    \\

\midrule
&$PGTG:$&\\
&$s^{ab}_{\;\;\mu}$ canonical spin density, $s_{ab}\equiv \delta \mathcal{L}_{m}/\delta \Gamma^{ab}$ &  $J=\frac{\partial \mathcal{L}_{m}}{\partial A}$  \\
Noether currents   &$\tau^{a}_{\;\mu}$ canonical energy-momentum density, $\tau_{a}\equiv \delta \mathcal{L}_{m}/\delta \vartheta^{a}$& \\
  &$MAG:$&(elec. charge, isospin,...)\\
(sources) &$\Delta^{ab}_{\;\;\mu}$ Hypermomentum, $\Delta_{ab}\equiv \delta \mathcal{L}_{m}/\delta \Gamma^{ab}$ & \\
 &$\tau^{a}_{\mu}$ canonical energy-momentum density, $\tau_{a}\equiv \delta \mathcal{L}_{m}/\delta \vartheta^{a}$& \\

\midrule

 &$PGTG:$& \\
 Excitations  &$H_{ab}=-\delta \mathcal{L}_{A}/\delta R^{ab}$& $H=-\delta \mathcal{L}_{A}/\delta F$  \\
 &$H_{a}=\delta \mathcal{L}_{A}/\delta T^{a}$& \\

\midrule

 &$PGTG:$& \\
Field equations  &$DH_{ab}-\varsigma_{ab}=s_{ab}$& $DH=J$  \\
 &$DH_{a}-\pi_{a}=\tau_{a}$& \\

\midrule
 &$PGTG:$& $dF=-F\wedge A$ \\
 Bianchi identities  &$dR^{a}_{b}+\Gamma^{a}_{c}\wedge R^{c}_{b}=-R^{a}_{c}\wedge\Gamma^{c}_{b}\quad(DR^{a}_{b}=0)$& ($DF=0$)\\
 &$DT^{a}=R^{a}_{c}\wedge\vartheta^{c}$& \\

\bottomrule

\end{tabular}

\label{table:III}
\end{table}

One of the most remarkable features of the gauge approach to gravity is the intimate link between group considerations and spacetime geometry. Non-rigid (local) spacetime symmetries require non-rigid (non-Euclidian) geometries. Moreover, as previously mentioned, by extending the symmetry group one is led to extend the spacetime geometry as well and, in this way, post-Riemann geometries have a natural place within gauge theories of gravity. For instance, while the translational gauge theories (TGTG) include non-vanishing torsion but zero curvature and non-metricity, Poincar\'{e} gauge gravity requires a RC geometry, and both Weyl(-Cartan) gauge gravity (WGTG) and conformal gauge gravity (CGTG) live on subsets of the more general metric-affine geometry with curvature, torsion and non-metricity. In particular, in WGTG the traceless part of the non-metricity vanishes. Both~the 10-parametric Poincar\'{e} group and the 11-parametric Weyl group are non-simple, meaning that they can be divided into two smaller groups (a non-trivial normal sub-group and the corresponding quotient group) and the natural extension from the corresponding theories of gravity into one with a simple group leads to CGTG. The 15-parametric conformal group $C(1,3)$ is simple (its only normal sub-groups are the trivial group and the group itself), but this extension requires a generalization of Kibble's gauge procedure, due to the fact that although locally $C(1,3)$ is isomorphic to $SO(2,4)$ its realization in $M_{4}$ (Minkowski spacetime) is non-linear \cite{Lord:1986ng}.

The PGTG can further be extended into the de Sitter or anti-de Sitter (A)dS gauge theories of gravity by localizing the $SO(1,4)$ or $SO (2,3)$ groups, respectively. Due to the fact that the (A)dS space is a maximally symmetric space which can be embedded into 5-dimensional Minkowski space (with~two or one time coordinates for AdS or dS, respectively), its isometries obey Lorentz type of algebra. Under a specific limit (by setting $l\rightarrow\infty$, where $l$ is a parameter of the group algebra), the~group goes into the Poincar\'{e} algebra. Depending on the choices of the Lagrangian, one can then have explicit \cite{MacDowell:1977jt} or spontaneous \cite{Stelle:1979aj} symmetry breaking from $SO(2,3)$ to $SO(1,3)$, for instance.

Another important class of extensions requires going beyond the Lie algebra by considering the algebra with anticommutators, in order to arrive at the super-Poincar\'{e} group \cite{Berkovits:1995ab} containing the usual Poincar\'{e} generators and proper supersymmetry (SUSY) transformations. These are generated by a Majorana spinor  which acts as the (anticommuting) generator of the transformations between fermions and bosons. The simple (with one supersymmetry generator) AdS supersymmetry generalizes the simple super-Poincar\'{e} algebra although it has the same generators, and it goes back to the super-Poincar\'{e} algebra under the same limit as the AdS group goes back into the PGTG. Further extensions include the consideration of a number $1<N<8$ of supersymmetry generators. The~gauging of these super-algebras lead directly to the bosonic gravity sector and therefore, supergravity (SUGRA) is an important class of supersymmetric gauge theories of gravity, extremely relevant for unification methods of bosons and fermion by the link it establishes between external (spacetime) symmetries and internal symmetries. In the self-consistent gauge approach, this class of theories needs to take into account post-Riemannian spacetime geometries, although many of the approaches have been done within the  Riemannian geometry \cite{Blagojevic:2012bc}.

To illustrate the structure of the gauge approach to gravity we next consider the PGTG in more~detail.

\subsection{The Gravity Yang-Mills Equations of Poincar\'e Gauge Theories of Gravity}

By applying to gravity a similar procedure as that of the Yang-Mills approach to gauge fields, one arrives at the mathematical structure of gauge theories of gravity. One starts with the (rigid) global symmetries of a matter Lagrangian with respect to a specific group of spacetime coordinate transformations, and the conserved Noether currents are identified. Then, by localizing (gauging) the symmetry group, the gauge gravitational potentials are introduced as well as the gauge covariant derivative and the respective field strengths, which are well known objects from differential geometry. Indeed, the gauge potentials represent the generators of the local symmetry group and couple to the respective conserved Noether currents, which act as sources of gravity. In practical terms, the~identification of the appropriate gauge field potentials comes from the requirement of covariance of $D\Psi$.

In PGTG the tetrads and the spin connection 1-forms are the gauge potentials, associated with translations, $T(4)$, and Lorentz rotations, $SO(1,3)$, respectively. Torsion and the curvature 2-forms are the respective field strengths. Torsion  can be decomposed into 3 irreducible parts $T^{a}=T^{a}_{(1)}+T^{a}_{(2)}+T^{a}_{(3)}$, made of a tensor part with 16 independent components, a vector part and an axial (pseudo) vector, both with 4 independent components\footnote{In the minimal coupling to fermions, only the axial vector torsion is involved.}. As for the curvature, it has
 36 independent components which can be decomposed into 6 irreducible parts: Weyl (10), Paircom (9), Ricsymf (9), Ricanti (6), scalar  (1), and pseudoscalar (1). In addition, there are 6 generators in the Lorentz group with 6 potentials ($\Gamma^{ab}=-\Gamma^{ba}$) and 6 spin (Noether) currents ($s^{ab}=-s^{ba}$). Analogously, there are 4 generators in the group of spacetime translations, which entail 4 gauge potentials $\theta^{a}$ and 4 conserved Noether currents $\tau^{a}$. By constructing the gravitational Lagrangian with the curvature and torsion invariants,  the potentials are coupled to the Noether currents via $24+16=40$ second order field equations.

Let us now build the different contributions of the matter and gravity fields to the PGTG. For~the former, we consider a matter Lagrangian $\mathcal{L}_{m}=\mathcal{L}_{m}(g_{ab},\theta^{c},D\Psi)$ such that the covariant derivative with respect to the RC connection, $D\Psi$, allows it to be invariant under local Poincar\'{e} spacetime transformations. There are two classes of Noether conserved currents: the canonical energy-momentum tensor density, which is equivalent to the dynamical tetrad energy-momentum density, $\tau_{a}\equiv \delta L_{m}/\delta \theta^{a}$; and the canonical spin density, which is equivalent to the dynamical spin density, $s_{ab}\equiv \delta L_{m}/\delta \Gamma^{ab}$. These currents couple to the gravitational potentials, acting as sources of gravity, and obey generalized conservation equations. As for the gravity sector in the action, it is constructed with the gauge-invariant gravitational field strengths in the kinetic part associated with the dynamics of the gravitational degrees of freedom.

The total Lagrangian density thus reads

\begin{equation} \label{eq:Lagpgtg}
\mathcal{L}=\mathcal{L}_{G}(g_{ab},\theta^{a},T^{a},R^{ab})+\mathcal{L}_{m}(g_{ab},\theta^{a}, D\Psi) \ .
\end{equation}

By varying this action with respect to the gauge fields of gravity ($\theta^{a},\Gamma^{ab}$) and the matter fields $\Psi$, we get the corresponding field equations. For the fermionic matter fields, the variational principle $\delta L_m/\delta \Psi=0$ leads to a generalization of the Dirac equation. As for the bosonic sector (gravity), the~inhomogeneous Yang-Mills equations in PGTG are

\begin{equation}
\label{inhomoPGTG}
DH_{a}-\pi_{a}= \tau_{a},\qquad D H_{ab}-\varsigma_{ab}= s_{ab} ,
\end{equation}
where $ H_{a}=-\partial \mathcal{L}_{G
}/\partial T^{a}$ and $ H_{ab}=-\partial \mathcal{L}_{G}/\partial R^{ab}$ are the 2-form excitations (field momenta) associated to torsion and curvature, and $\tau_{a}\equiv \delta \mathcal{L}_{m}/\delta \theta^{a}$ and $s_{ab}\equiv \delta \mathcal{L}_{m}/\delta \Gamma^{ab}$ are the 3-form canonical energy-momentum and spin currents. The 3-forms $\pi_{a}$ and $\varsigma_{ab}$ can be interpreted as the energy-momentum and spin of the gravitational gauge fields, respectively, defined as

\begin{equation}
\pi_{a}\equiv  e_{a}\iprod L_{G}+(e_{a}\iprod T^{b})\wedge H_{b}+(e_{a}\iprod R^{cd})\wedge H_{cd} ,
\end{equation}
\begin{equation}
\varsigma_{ab}\equiv-\theta_{[a}\wedge H_{b]} \ .
\end{equation}

For a given theory, one only needs to compute the excitations, the source currents and the gravitational currents from the Lagrangian density (\ref{eq:Lagpgtg}) and substitute directly in the inhomogeneous equations (\ref{inhomoPGTG}).
Note that in this formalism of exterior forms these field equations are completely metric-free, fully general and coordinate-free, with well defined gravitational energy-momentum and spin currents. The PGTG have two sets of Bianchi identities, previously introduced as $D R^{a}_{b}=0$ and $DT^{c}= R^{c}_{d}\wedge\theta^{d}$, which are intrinsic to the geometrical structure of RC spacetime. Via the field equations, these can be related to the generalized conservation equations for the energy-momentum and the spin currents.

Regarding the constitutive relations in these theories, the 2-form excitations expressed in terms of the field strengths (torsion and curvature), $H_{a}=H_{a}(T^{c},R^{bd})$ and $H_{ab}=H_{ab}(T^{c},R^{bd})$,
represent two sets of constitutive relations and are implicit in the Lagrangian formulation. These excitations are in exact analogy to the canonically conjugate momenta of classical mechanics, while torsion and curvature are the generalized velocities for the gravitational degrees of freedom represented by the gauge potentials. These constitutive relations in vacuum can be interpreted as  describing the gravitational propagation properties of the spacetime physical manifold. It is via these relations that the conformally  invariant part of the metric is introduced, via the Hodge star operator for instance. Moreover, in~these relations the coupling constants of the theory are required in order to adequately convert the dimensions of the field strengths (field velocities) to the excitations (field momenta). As constitutive relations for the spacetime vacuum itself, one can postulate that such coupling constants characterize physical properties of the spacetime manifold endowed with gravitational geometrodynamics. This~hypothesis is in clear analogy to the similar interpretation for the electromagnetic properties entering in the corresponding constitutive relations (see Reference \cite{Cabral:2016yxh} for further details).

\subsubsection{Quadratic Poincar\'{e} Gauge Gravity}

PGTG with Lagrangians quadratic in the curvature and torsion invariants have been investigated within cosmology, gravitational waves, and spherical solutions, see for example, References \cite{Obukhov:2018bmf,Obukhov:2017pxa,Blagojevic:2018dpz,Blagojevic:2015zma,Cembranos:2019mcb}. The PGTG class is fundamental given the importance of the Poincar\'{e} symmetries in relativistic field theories, and the most general quadratic Lagrangian ({\it $\grave{a}$ la} Yang-Mills) contains parity breaking terms induced by the richer RC geometry with curvature and torsion \cite{Obukhov:2018bmf}. It can be written as
 \begin{eqnarray}
 L&=&\dfrac{1}{2\kappa^{2}}\Big[ (  a_{0}\eta_{ab}+\bar{a_{0}}\theta_{a}\wedge\theta_{b}) \wedge R^{ab}-2\Lambda\eta 
 -T^{a}\wedge \sum_{I=1}^{3}\left( a_{I}( \star T_{a}^{(I)})+\bar{a}_{I} T_{a}^{(I)}\right) \Big]  \nonumber \\
 &&-\dfrac{1}{2\rho}R^{ab}\wedge \sum_{I=1}^{6}\left(b_{I}(\star R_{ab}^{(I)})+\bar{b}_{I} R_{ab}^{(I)} \right).
 \label{quadraticpgtg}
 \end{eqnarray}

The first term in the first line corresponds to the ECSK theory plus the Holst term $\sim (\theta_{a}\wedge\theta_{b}) \wedge R^{ab} $, where $\eta_{ab}\equiv e_{b}\iprod\eta_{a}=\star(\theta_{a}\wedge\theta_{b})$\footnote{Here $\eta_{a}=e_{a}\iprod\eta=\star\theta_{a}$ is a 3-form and $\eta=\star 1$ is the natural volume 4-form.}. The second term corresponds to a cosmological constant. The~second line contains the terms quadratic in the torsion field strength and the index $I=1,2,3$ runs over the three irreducible pieces of torsion. In the third line we have the curvature quadratic terms and  the index $I=1,...,6$ runs over the six irreducible pieces of curvature. The free parameters include the $2+6+12=20$ ($a,b$) coefficients, plus the  cosmological constant and the sometimes called ``strong gravity'' parameter $\rho$. In this Lagrangian all terms with the coefficients with a bar $\bar{a_{0}}, \bar{a}_{I}, \bar{b}_{I}$ break the symmetry under parity transformations. For specific choices and assumptions this Lagrangian includes, for instance, GR itself, the teleparallel equivalent to GR, or the ECSK theory. In the latter, Dirac fermions have axial-axial contact interactions (the Hehl-Data term) with a repulsive character, while in general quadratic PGTG this contact spin-spin interaction is generalized by predicting a propagating interaction. In particular, intermediating  gauge bosons with spins $s=0,1,2$ are predicted, which correspond to massive or massless scalar, vector\footnote{This is actually a torsion axial vector which couples to elementary particles.} and tensor propagating modes, respectively. In these GW fields there are odd parity (parity breaking) modes which could manifest themselves as signatures of chirality in the GW cosmological backgrounds from the early Universe\footnote{For that, one needs non-planar detectors, 3-point correlation functions analysis and sufficient signal/noise ratio, besides a clear distinction from other possible GW sources with a similar power spectrum signature. }. Let us also note that in PGTG it is possible to identify ghost-free Lagrangians which can also be quantized \cite{Sezgin:1979zf,Kuhfuss:1986rb}.

\subsubsection{The Teleparallel Equivalent of GR}

Choosing the Weitzenb\"{o}ck spacetime geometry for PGTG, which implies keeping only torsion while setting both curvature and non-metricity to zero, it is possible to formulate a  Lagrangian quadratic in torsion subject to some specific restrictions  (such as an appropriate choice for the Weitzenb\"{o}ck connection). This theory can be formulated from a translational gauge theory of gravity perspective. Heuristically, one gets in this case

\begin{equation}
D_{\alpha}T^{\;\;\alpha\gamma}_{c}+(...) \sim \kappa^{2}\tau_{c}^{\;\;\gamma} \ ,
\end{equation}
or, in terms of the tetrads
\begin{equation}
\Box \theta^{a}_{\;\mu}+(...) \sim \kappa^{2}\tau^{a}_{\;\mu} \ ,
\end{equation}
where the missing terms on the left-hand side are non-linear terms, and $\Box$ is a d'Alembertian operator. The last equation resembles its GR counterpart, $\Box g_{\mu\nu}+...\sim \kappa^{2}T_{\mu\nu}$, and it turns out that both equations yield the same gravitational phenomenology  for matter described by fundamental scalar or Maxwell fields, where the canonical $\tau_{\mu\nu}$ and the dynamical (Einstein-Hilbert) $T_{\mu\nu}$ energy-momentum tensors coincide. For fermions the theories are not fully equivalent. It is interesting  to point out that the quadratic (in torsion) Lagrangian of this teleparallel equivalent of GR (TEGR) is locally Lorentz invariant and equivalent to the Hilbert-Einstein Lagrangian \cite{BeltranJimenez:2019tjy}, but if one wants to formulate GR as a gauge theory then one must gauge the translational group instead of the Lorentz group. This~provides yet another motivation to go beyond GR using in this case the teleparallel formulation of gravity to explore exact solutions, post-Newtonian limit, gravitational waves, and so forth, References \cite{Jimenez:2019ghw,Hohmann:2018jso,Bahamonde:2017wwk,Bahamonde:2020cfv,delaCruz-Dombriz:2018nvt}, since it is plausible to consider the whole Poincar\'{e} symmetries in Nature to be valid, not only the translational group.

\subsubsection{Einstein-Cartan-Sciama-Kibble Gravity}

In the formalism of exterior forms, the ECSK Lagrangian can be written as

\begin{equation}
\mathcal{L}=\dfrac{1}{2\kappa^{2}}\eta_{ab}\wedge R^{ab} \ .
\end{equation}

The field equations are then obtained by varying this action with respect to the tetrads and the (Lorentzian) spin connection. In the more common tensor formalism this Lagrangian corresponds to the linear Lagrangian in the curvature scalar, yielding the action

\begin{equation} \label{eq:actionEC}
S_{\rm EC}=\dfrac{1}{2\kappa^2} \int d^4x \sqrt{-g}R(\Gamma) + \int d^4x \sqrt{-g}\,\mathcal{L}_{m}\,.
\end{equation}

In this expression $\kappa^2=8\pi G$ with $G$  Newton's coupling constant, and $g$ is the determinant of the spacetime metric $g_{\mu\nu}$. In the RC spacetime the curvature scalar $R(\Gamma)$ defined via (\ref{ricciRC}) includes terms quadratic in torsion, while the matter Lagrangian, $\mathcal{L}_m=\mathcal{L}_{m}(g_{\mu\nu},\Gamma,\Psi_m)$ depends on the metric and the matter fields, $\Psi_m$, and also on the contortion (i.e., on torsion) via the covariant derivatives.

The Cartan equations can be obtained by varying the  action (\ref{eq:actionEC}) with respect to the contortion tensor $K^{\alpha}_{\;\beta\gamma}$ (or, alternatively, with respect to the spin connection), and the corresponding result can be written as

\begin{equation}
T^{\alpha}_{\;\beta\gamma}+T_{\gamma}\delta^{\alpha}_{\beta}-T_{\beta}\delta^{\alpha}_{\gamma}=\kappa^2 s^{\alpha}_{\;\beta\gamma} \ ,\label{cartaneqs}
\end{equation}
where
\begin{equation}
\label{spintensor}
s^{\gamma\alpha\beta}\equiv\dfrac{\delta \mathcal{L}_{m}}{\delta K_{\alpha\beta\gamma}} \ ,
\end{equation}
is the spin density tensor, while $T_{\beta}\equiv T^{\gamma}_{\;\beta\gamma}$ is the torsion (trace) vector. Cartan's equations (\ref{cartaneqs}) imply that torsion is related to the spin density of matter fields via linear and algebraic relations and, therefore, in the absence of spin (as in vacuum) torsion vanishes.
On the other hand, the variation of the action~(\ref{eq:actionEC}) with respect to the spacetime metric $g_{\mu\nu}$ (or, alternatively, with respect to the tetrads) yields the generalized Einstein equations, which,can be written as $G_{\mu\nu}=\kappa^2 \tau_{\mu\nu}$, where $G_{\mu\nu}$ is the Einstein tensor of the RC geometry and $\tau_{\mu\nu}$ is the canonical energy-momentum. These equations can also be conveniently written as

\begin{equation}
\tilde{G}_{\mu\nu}=\kappa^2 T^{\rm eff}_{\mu\nu} \ ,\label{ECeqs}
\end{equation}
where $\tilde{G}_{\mu\nu}$ is the Einstein tensor computed with the Levi-Civita connection, while the effective stress-energy tensor

\begin{equation}
T^{\rm eff}_{\mu\nu}=T_{\mu\nu}+U_{\mu\nu}=-\frac{2}{\sqrt{-g}} \frac{\delta (\sqrt{-g}\mathcal{L}^{\rm eff}_{m})}{\delta g^{\mu\nu}}\,,
\end{equation}
includes the (dynamical) metric energy-momentum tensor of the matter fields, $T_{\mu\nu}=-\tfrac{2}{\sqrt{-g}} \tfrac{\delta (\sqrt{-g}\mathcal{L}_m)}{\delta g^{\mu\nu}}$, and the additional piece $U_{\mu\nu}=-\tfrac{2}{\sqrt{-g}} \tfrac{\delta (\sqrt{-g}C)}{\delta g^{\mu\nu}}$, with $C\equiv-\tfrac{1}{2\kappa^2}
\Big(K^{\gamma}_{\;\beta\gamma}K^{\alpha\beta}_{\;\;\;\alpha}+K^{\alpha\lambda\beta}K_{\lambda\beta\alpha}\Big)$, which contains corrections which are quadratic in torsion $U\sim\kappa^{-2}T^{2}$. This piece can also be expressed in terms of the spin tensor using Cartan's equations (\ref{cartaneqs}), i.e., $U\sim \kappa^2 s^2$. Note also that, in general, torsion also contributes to the energy-momentum tensor $\tau_{\mu\nu}$, since the covariant derivatives present in the kinetic part of $\mathcal{L}_m$ introduce new terms depending on torsion via either minimal or non-minimal couplings.
Since $U\sim \kappa^2 s^2$, Equation~(\ref{ECeqs}) defines a typical density, known as Cartan's density, and given by $\rho_{C} \sim 10^{54}$g/cm$^3$ (if one considers nuclear matter\footnote{In cosmological applications the critical density can be written as $\rho_{\rm crit}\sim m/\lambda_{\rm Comp}l^{2}_{\rm Pl}$, where $l_{\rm Pl}$ and $\lambda_{\rm Comp}$ are Planck's length and Compton wavelength, respectively. For electrons we get $\rho_{\rm crit}\sim 10^{52}$g/cm$^3$, corresponding to $T_{\rm crit}\sim 10^{24}K$ and around $t \sim 10^{-34}$s after the Big Bang.}). Therefore, in principle, ECSK theory can only introduce significant physical effects in environments of very large spin densities, which might arise in the early universe or in the innermost regions of black holes.

To conclude this part, let us mention that, once the coupling to fermions is considered, the~generalized Dirac (Hehl-Data) equation for spinors in RC spacetime, coupled to the ECSK gravity, is cubic in the spinors and includes a torsion induced spin-spin (axial-axial) contact interaction. This~type of interactions have been searched for in particle physics including studies at HERA, LEP and Tevatron in electron-proton scattering \cite{Goy:2004ap,Adloff:2003jm}.

\subsection{Quadratic Gauge Gravity Models in Metric-Affine Gravity}

One can generalize the PGTG by considering the affine group as the gauge symmetry group for gravity. If this is performed {\it $\grave{a}$ la} la Yang-Mills, then one gets quadratic models as in quadratic PGTG, where the quadratic terms are sometimes referred to as (hypothetical) ``strong gravity'' terms. This idea has been recovered from time to time, as in Yang \cite{Yang:1974kj}, in the tensor dominance model \cite{Isham:1971gm} or in
 chromogravity \cite{Neeman:1995rjn}. Depending on the choice of the Lagrangian, the strong gravity (bosonic sector) can be very massive or massless. In some respect these gauge bosonic gravity fields are similar to the Yang-Mills bosons, and if they are massive it is typically assumed that the masses are of the order of the Planck mass or even above. As in quadratic PGTG, in quadratic metric-affine models there are intermediate gauge bosons with spins ($s=0,1,2$) corresponding to scalar, vector or tensor modes (massive or massless). In this respect, the quadratic model in Equation~(\ref{quadraticpgtg}) for the PGTG can be extended to metric-affine gravity by including terms with $Q_{ab}$ and $R_{(ab)}$ which are both zero in PGTG, in accordance with the choice of a Lorentzian connection and, therefore, of zero non-metricity (\mbox{see References \cite{Hehl:1976my,Lord:1978qz,Hehl:1989ij}}).

As an example of a quadratic metric-affine model, one can consider a Lagrangian density (for the bosonic-gravity sector) of the form (schematically)

\begin{equation}
\mathcal{L}_{MAG}\sim \dfrac{1}{\kappa^{2}}\left( R+T^{2}+TQ+Q^{2}\right)+\dfrac{1}{\rho}\left( W^{2}+Z^{2}\right) ,
\end{equation}
where $W_{ab}\equiv R_{[ab]}$, and $Z_{ab}\equiv R_{(ab)}$ are known as the ``rotational curvature'' and the ``strain curvature'', respectively.
The terms proportional to the coupling constant $\rho$ are referred to as strong gravity terms, in contrast to the terms that are proportional to the ``weak'' gravity coupling constant $\kappa^{2}$. Note that in this model the connection is non-Lorentzian and the non-metricity 1-form (represented by $Q$) is non-vanishing. The bosonic sector of metric-affine gravity was analyzed, for instance, \mbox{in~References \cite{Hehl:1998nx,Hehl:1976my,Lord:1978qz,Hehl:1989ij},} while the fermionic part is more delicate (see for instance References \cite{Neeman:1978kyt,Neeman:1985qoa,Neeman:1985bip,Hehl:1994ue}). In this respect, there is no finite dimensional spinor representation of the $GL(4,\Re)$ group, which~leads to the introduction of the ``world spinors'' (with infinite components) of Ne'eman, and to the corresponding generalization of Dirac equation. The world spinor formalism is related to Regge trajectories, which are themselves related to spin-2 excitations of hadrons (see Reference \cite{Sijacki:2005fe}).

For an interesting review on exact solutions in MAG see Reference \cite{Hehl:1999sb}. Further research in this context involves cosmological scenarios \cite{Puetzfeld:2004yg}, and one should also mention exact spherically symmetric solutions with $Q\sim 1/r^{d}$ \cite{Obukhov:1996pf}, which suggests the existence of massless modes.

\subsection{Probing Non-Riemannian Geometry with Test Matter}

It is known that torsion can give rise to precession effects in systems with intrinsic spin, for~example, elementary particles such as the electron, or baryons such as the neutron \cite{Rumpf:1979vh,Audretsch:1981xn}. This is a model-independent result which can be obtained from a (WKB) semi-classical approximation of the Dirac fermionic dynamics in a RC spacetime. In principle this prediction can be used to  distinguish between the spacetime paradigms of GR and TEGR. If $v$ is the polarization vector of the (intrinsic) spin then one can deduce the simple expression

\begin{equation}
\dot{v}=3\breve{t}v \ ,
\end{equation}
where the axial vector $\breve{t}$ is given by $\breve{t}^{\alpha}\equiv-\epsilon^{\alpha\beta\gamma\delta}T_{\beta\gamma\delta}$. In this sense it is plausible that experiments similar to the Gravity Probe-B but using gyroscopes with intrinsic (macroscopic) spin can be used to constrain or detect the effects of the (hypothetical)  torsion around the Earth \cite{Ni:2004jj}. There have been quite a number of studies on spin precession effects induced by torsion (see also Reference \cite{Nitsch:1979qn}). On the other hand, Lammerzahl have set experimental limits for detecting torsion ($\vert T\vert\sim 10^{-15} m^{-1}$) using Hughes-Drever (spectroscopic) type of experiments \cite{Lammerzahl:1997wk}.

Moreover, one can predict torsion effects on the energy levels of quantum systems. From the Dirac Lagrangian of a fermion minimally coupled to the background RC geometry

\begin{equation}\label{eq:mattfer}
\mathcal{L}_{\rm Dirac}=\dfrac{i\hbar}{2}\left(\bar{\psi}\gamma^{\mu}D_{\mu}\psi-(D_{\mu}\bar{\psi})\gamma^{\mu}\psi\right)-m\bar{\psi}\psi \ ,
\end{equation}
(for spinors $\psi$ and their adjoints $\bar{\psi}$) one deduces the Dirac equation

\begin{equation}
\label{eq:DiracFockIvaneckoHeisenberg}
i\hbar \gamma^{\mu}\tilde{D}_{\mu}\psi-m\psi =
-\dfrac{3\hbar}{2} \breve{T}^{\lambda}\gamma_{\lambda}\gamma^{5}\psi \ ,
\end{equation}
which can be analyzed in the flatness limit. If we assume a static axial torsion vector $\breve{T}^{\lambda}$ along the $x^{3}$-direction, then we get the time-independent wave equation

\begin{equation}
\label{eq:Diracflatlimittimeind}
-i\hbar \gamma^{k}\partial_{k}\psi+\Big(m
-\dfrac{3\hbar}{2} \breve{T}^{3}\gamma_{3}\gamma^{5}\Big)\psi(\vec{r})=\gamma^{0}E\psi(\vec{r}) \ .
\end{equation}

From this equation one obtains two possible values for the energy levels, depending on whether the fermionic spin is aligned or anti-aligned with respect to the background (axial) torsion, that is \cite{FCLR20},

\begin{equation}
E^{2}=p^{2}+\left(m\pm\dfrac{3\hbar}{2}\breve{T}^{3}\right)^{2}.
\end{equation}

Similarly, if non-minimal couplings between the torsion trace vector and Dirac axial vector and/or between torsion axial vector and Dirac vector $\bar{\psi}\gamma^{\mu}\psi$, are present, then the parity symmetry is broken and the corresponding energy level corrections due to torsion will contain the signatures of those parity breaking interactions. Therefore, tests with advanced spectrographs might be able to probe torsion effects on quantum systems. Also, the previously mentioned spin-spin contact interactions of the ECSK theory, or the propagating spin-spin interactions mediated by gravitational gauge ($s = 0,1,2$) bosons might be tested/constrained in laboratory experiments and cosmological GW probes.

Regarding non-metricity, the gauge approach to gravity clearly shows that the hypermomentum currents, such as the dilatations or the shear currents, couple to the trace vector part and the shear part of the non-metricity, respectively. This coupling is evident since, as we have seen, the~connection couples to hypermomentum and a non-Lorentzian connection implies non-metricity. Further developments have shown that, if torsion can be measured by the spin precession of test matter with intrinsic spin, then the non-metricity of spacetime can be measured by pulsations (mass~quadrupole excitations) of test matter with (non-trivial) hypermomentum currents. In order to be ``sensitive'' to non-Riemann geometries, test matter should carry dilatation, shear, or spin currents, whether macroscopic or at the level of fundamental fields/particles. In the latter case,  the Regge trajectories provide an adequate mathematical illustration of test matter as Ne'eman's world spinors with shear.

Let us also point out that Obukhov and Puetzfeld \cite{Puetzfeld:2007ye} have derived the equation of motion for matter fields in metric-affine gravity. By making use of the Bianchi identities one can arrive at the following expression for the translational Noether current:

\begin{equation}
\tilde{D}\Big[\tau_{a}+\Delta^{bc}\left(e_{a}\iprod N_{bc} \right)\Big]+\Delta^{bc}\wedge\left(\pounds_{e_{a}}N_{bc} \right)=s^{bc}\wedge(e_{a}\iprod\tilde{R_{cb}}),
\end{equation}
where, as usual, the tilde refers to quantities defined in the Riemann geometry, while $\tau_{a}$, $\Delta^{bc}$, $s^{bc}$ are the canonical energy-momentum, hypermomentum current, and spin current, respectively, \mbox{and~$N_{bc}=\Gamma_{bc}-\tilde{\Gamma}_{bc}$} gives the non-Riemannian piece of the connection 1-form, that is, the distortion 1-form. Note that in the right-hand side of this equation we can identify the Mathisson-Papapetrou force density for matter with spin. In standard GR, one obtains $\tilde{D}\tau_{a}=0$, which gives the geodesic equation for spinless matter with energy-momentum, while for $N_{bc}=0$ we get \mbox{$\tilde{D}\tau_{a}=s^{bc}\wedge(e_{a}\iprod\tilde{R}_{cb})$}, which is the Mathisson-Papapetrou equation in GR for matter with spin.  In this equation, if~matter has neither (intrinsic) spin, nor dilatation/shear currents, then it follows the Riemannian (extremal~length) geodesics, regardless of the geometry of spacetime or of the form of the Lagrangian in metric-affine~geometry.

\subsection{Metric-Affine Geometry and Lorentz Symmetry Breaking}

Regarding the analogous law for the $GL(4,\Re)$ Noether current we have
\begin{equation}
D\Delta^{a}_{\;b}+\theta^{a}\wedge\tau_{b}=\tau^{a}_{\;b} \ .
\end{equation}

In PGTG the connection is Lorentzian $\Gamma^{ab}=-\Gamma^{ba}$, while in the WGTG the trace part of the connection $\tfrac{1}{4}g^{ab}\Gamma^{c}_{\;c}$ is also non-vanishing. A fully independent connection in metric-affine theory is given by the expression in (\ref{independentconnection}). It is this linear connection that couples to the  (intrinsic) hypermomentum current (see Equation~(\ref{eq:hypermomentum})). In particular, the  Lorentzian connection couples to spin $s_{ab}$ carrying $SO(1,3)$ charges, while the trace part couples to the dilatation current $\Delta^{c}_{\;c}$, and the shear part of the connection (see Equation~(\ref{MAGconnection})) couples to the shear current $\bar{\Delta}_{ab}$ that carries $SL(4,\Re)/SO(1,3)$ (intrinsic) shear charges. The shear current seems to be related to the Regge trajectories \cite{Hehl:1994ue} and these represent spin-2 excitations of hadrons with the same internal quantum numbers. The relation between the Regge trajectories (which can be described by the group $SL(3,\Re)$ and this can be embedded in the $SL(4,\Re)$ one) and the hypermomentum shear charges, remains an open question under study and its validity seems to point to a quite remarkable and promising connection between the strong interaction of hadrons and spacetime post-Riemann geometry. The shear charge is actually a measure of the breaking of Lorentz invariance. The bottom line is that, in order to get Lorentz symmetry breaking, one does not require the introduction of extra particle degrees of freedom, but this can be obtained solely via the geometrical structures of spacetime, namely a non-Lorentzian connection. The presence of a non-Lorentzian connection implies the non-vanishing non-metricity and the non-vanishing strain curvature $R_{(ab)}$.

\subsection{A Word on the Formulations of GR}

The (canonical) metric formulation of GR requires a pseudo-Riemannian manifold with a symmetric, $\Gamma^{\alpha}_{\mu\nu}=\Gamma^{\alpha}_{\nu\mu}$, and metric-compatible, $\nabla_{\alpha}^{\Gamma} g_{\mu\nu}=0$, affine connection (the Levi-Civita one).  However, it is well known nowadays that the teleparallel equivalent of GR, formulated in the Weitzenb\"{o}ck spacetime, yields a dynamically equivalent theory to metric GR, with exactly the same predictions \cite{BeltranJimenez:2019tjy,Hayashi:1979qx,Aldrovandi:2013wha,Maluf:2013gaa}. Besides this approach from a translational gauge principle under specific assumptions, there is yet another formulation equivalent to GR based on zero curvature and torsion but non-zero non-metricity, called symmetric teleparallel gravity, whose properties have begun to be unravelled very recently  \cite{BeltranJimenez:2017tkd,BeltranJimenez:2018vdo,Jarv:2018bgs,Iosifidis:2018zjj,Harko:2018gxr}.

Since these are, to some extent, equivalent gravitational models under different spacetime paradigms, one may ask if there is any guiding principle which could determine what spacetime geometry and degrees of freedom can represent gravity at its most fundamental level. The application of the gauge approach to gravity shows clearly that GR can be formulated as a translational gauge theory and, therefore, lives on a subset of the RC spacetime. On the other hand the generalization of the translational symmetry to the Poincar\'{e} symmetries points towards the direction of the PGTG formulated in the RC spacetime geometry. It is relevant to underline (once again) that the three mentioned approaches to GR have different assumptions regarding the spacetime geometrical paradigm, but their equivalence breaks as soon as one considers  matter with spin (spinors). Indeed, if~one could measure the different effects of non-Riemannian geometries upon matter, one might be able to distinguish between these spacetime paradigms.

\section{Discussion and Future Outlook\label{Conclusion}}
\unskip

\subsection{Spacetime Paradigms}

In this work we explored geometrical methods (post-Riemann spacetime geometries and Cartan's exterior calculus of forms) and symmetry principles in the gauge approach to gravity, and how these topics might point towards a new perspective over the spacetime paradigm. We also briefly considered the pre-metric formulation of classical electrodynamics. In this broad perspective, the~conformal-causal structure is argued to be more fundamental than the metric structure (the primacy of the conformal geometry) and the absoluteness of the spacetime metric is then abandoned at the fundamental level. We~also established the analogies between the pre-metric canonical formulation of gauge theories of gravity and the pre-metric equations and mathematical objects of general Yang-Mills fields. The~theoretical formulation of the Lagrangians in these theories (gravity and Yang-Mills) implicitly presuppose the assumption of the specific form for the so-called constitutive relations between the field strengths (the generalized field velocities) and the excitations (conjugate field momenta). These~relations can be interpreted as constitutive relations for the spacetime itself. Moreover, in this discussion we highlighted the hypothesis that the physical constants or coupling parameters that enter in such relations reflect physical properties of the spacetime manifold, itself regarded as a mathematical object that represents a truly physical system.

By endowing the classical spacetime with physical properties, the concept of classical vacuum with properties such as electric permitivity, magnetic permeability, and so forth, becomes somehow disposable or simply dual to the very notion of a physical spacetime. Moreover, the properties of this physical spacetime might change from point to point and this scenario fits well within the scalar-tensor (Brans-Dicke, etc.), vector-tensor or tensor-tensor extensions to GR. The idea that these properties of spacetime can be described by fields can have implications to spacetime symmetry considerations, i.e., the invariance of the physics under groups of spacetime coordinate transformations, and also a link to the Mach's ideas and the breaking of spacetime (metric) absoluteness. Therefore, this scenario of a physical spacetime with non-Riemann geometry and physical properties described by fields fits naturally very well in the assumption of the primacy of the conformal-causal structure. This means that the so-called constants can change from place to place in space (non-homogeneity) and with spatial direction (anisotropy) and still preserve the local conformal-causal symmetry and, therefore, by extension the causal structure.  Consequently, what is assumed to be physically relevant  are the conformally-invariant properties. In other words, one must seek for an extended spacetime manifold with  post-Riemann geometries, physical properties, and a fundamental conformal symmetry.

Keeping this trail of thought, one arrives at the conclusion that the intrinsic physical properties of spacetime include energy-momentum, hypermomentum (including spin), electromagnetic and thermodynamical properties associated to gravity and to the spacetime. The latter actually suggests the existence of microscopic degrees of freedom, and if these are assumed to constitute a numerable set, then a consistent picture should go beyond the classical manifold and incorporate some degree of discreteness or gravity quantization. Therefore, a physical spacetime manifold with locally invariant conformal-casual structure, with intrinsic physical properties (electromagnetic, thermodynamical, etc.) and possibly a quantum nature seems the natural hypothesis to address the unification of spacetime, matter fields and the quantum and classical vacuum.

\subsection{Perspectives on Unification Methods in Fundamental Field Theories}
Let us now explore in more detail the hypothesis stated above, as well as its relation to various aspects of unification in physics. We shall put here our emphasis upon the conceptual background motivated by mathematical considerations and, in particular, from geometrical methods and symmetry~principles.

Firstly, the geometrical methods in the pre-metric formalism of electromagnetism, Yang-Mills and gravity field equations, using the calculus of exterior forms, together with the corresponding constitutive relations (that can be interpreted as (spacetime) constitutive relations, suggest the following: (i) the primacy of the conformal-causal structure (the conformally-invariant part of the metric) over the full metric structure, and therefore, (ii) the assumption of absoluteness of the spacetime metric (``absolute spacetime'') is abandoned at the fundamental level. Also one postulates that the fundamental coupling constants entering in the (vacuum) constitutive relations represent physical properties of spacetime, not necessarily spatially homogeneous and isotropic (constants), while~respecting  the local conformal symmetries. As a consequence we emphasize again here the first unification: the identification of the physical classical vacuum with physical classical spacetime.

Secondly, from gauge symmetries in gravitation and post-Riemann geometries we find the need to consider a classical spacetime with general metric-affine geometry, namely, having non-vanishing curvature, torsion and non-metricity. Indeed, the gauge approach to gravity  requires the existence of post-Riemann geometries associated to gravitational phenomena. This motivates the search for its signatures via astrophysical and cosmological effects, including GW probes and effects in particle physics. On the other hand, the idea that the spacetime metric might not be fundamental is rooted on the mathematical consistency of gauge theories of gravity which links  the symmetries of physics under  coordinate transformations to the geometrical paradigm of spacetime. The theory identifies clearly, for~each specific gauge group, the underlying spacetime geometry, the Noether currents (which are the sources of the gravitational field) and the gauge potentials (geometrical degrees of freedom)  and their field strengths. In this formalism the spacetime metric is not a fundamental field. This is clearly seen in the language of forms, where an explicit pre-metric approach to the field equations is completely general, coordinate-free and covariant.

 The metric structure can be assumed \textit{a posteriori}, in particular, via the  constitutive relations (which relate the field strengths or field velocities, that is, curvature and torsion, to the canonically conjugate field momenta). More specifically, using these relations, via the Hodge star operation, only the conformally invariant part of the metric structure is involved in the coupling between the field strengths and the Noether sources, rather than the full metric structure. Therefore, symmetry principles and geometrical methods in gauge theories of gravity suggest again that the (full) metric structure is not fundamental, and in models with larger symmetry groups beyond the Poincar\'{e} group the paradigm of spacetime (metric) absoluteness is not valid, that is, the metric changes under specific coordinate transformations between local observers. If non-metricity is present, the connection is non-Lorentzian and, as a consequence, one gets the breaking of Lorentz symmetry. It is this type of geometry that is associated to the conformal structure that emerges from linear electrodynamics \cite{Obukhov:1999ug}.

 Regarding unified symmetry groups and symmetry breaking, it is known that general symmetry groups can be broken into smaller groups within phase transitions. In principle, these are expected to have occurred in the early Universe in clear analogy to the standard model (and beyond the standard model) of interactions, leading to first order phase transitions and the generation of GW emission in the form of a stochastic GW background. This GW signature of cosmological origin (which~might include polarization features due to parity breaking gravitational physics of general quadratic Yang-Mills gauge theories of gravity) might be detectable by future missions such as LISA~\cite{Barausse:2020rsu}.  Of particular interest is the conformal group of CGTG, that may be broken into the Poincar\'{e} group. This necessarily includes the breaking of scale invariance and the emergence of natural physical scales and the corresponding (Lorentz invariant) fundamental constants. This presupposes the existence of a scale-invariant cosmological epoch where the properties of the physical spacetime obey perfect conformal symmetry and its geometry, and also a transition into a broken symmetry phase where the spacetime metric appears to have a (local) absolute nature. Whether there might be other extreme physical regimes (such as the innermost regions of ultra-compact objects and black holes) where scale invariance, or even the full conformal symmetry, is recovered, remains an open~question.

 Another suggestion is that curvature, torsion and non-metricity might be inter-convertible.  This seems appropriate from the point of view of the generalized Bianchi identities, \mbox{(\ref{BianciRC}) and (\ref{bianchinonmetricity}),} of metric-affine geometries relating the field strengths of gauge theories of gravity (curvature, torsion and non-metricity). This is somewhat analogous  to the magnetic-electric inter-conversion of electromagnetism.  Recall that the $dF=0$ is a Bianchi identity giving magnetic flux conservation and Faraday law. These post-Riemann Bianchi identities are implicit to the spacetime geometrical structures and express some sort of gravitational flux conservation, and are compatible with the Noether currents conservations of the gauge approach to gravity. These relations point towards this notion of inter-conversion, which could open new avenues for the study of gravitational phenomena in extreme regimes.

Moreover, this hypothesis resembles the Weyl-Ricci conversion (the Weyl conjecture) in cosmology, within GR. According to it, the Ricci curvature dominates completely the very early Universe and the Weyl curvature dominates the late-Universe, in such a way that  the Ricci part of the Riemann tensor is converted (transformed) into the Weyl part  as the Universe expands, and forms gravitationally bound structures, asymptotically dominated by black holes.  Similarly, the dynamical transformations (conversions) between the three parts inside the full curvature of MAG  can be compatible to the (generalized) Bianchi identities involving the full curvature.   This interpretation is further reinforced by the existence of formal maps between ``equivalent'' descriptions of GR in the spacetimes with torsion, curvature or non-metricity alone (and also between the generalizations of GR in the respective spacetime paradigms). Although under specific formal conditions the phenomenology of these theories might be equivalent, the spacetime paradigms are obviously different and the minimal/non-minimal couplings of fermionic fields to these geometries can break the equivalences. The idea that the Poincar\'e or the $GL(4,\Re)$ groups  should be fundamental points towards generalizations of GR in any of these quasi-equivalent descriptions, via PGTG or MAG, and in this context the mathematical structure of the gauge approach seems to be compatible with the interpretation that curvature, torsion and non-metricity can indeed be inter-convertible.

In addition to the topics discussed above it is worth mentioning that the gauge approach is transversal to both gravity and Yang-Mills fields of the standard model of particles and interactions. Although in the first case one speaks of external or spacetime symmetries, while in the second case the spinorial character of elementary particles/fields requires internal symmetries, in both cases there is a deep relation between group theory (symmetries) and the dynamics of fundamental fields via geometrical methods. The gauge formalism thus highlights its potential for bridging spacetime geometry  and gravity with the other fundamental interactions. The challenges are precisely those of how to relate internal spaces and its symmetries with the spacetime geometry and its symmetries in a unified way. In this context, we mentioned briefly the supersymmetric gauge theories of gravity, which somehow aim to achieve that goal by enlarging the gauge formalism and unifying bosons and fermions under the same mathematical structure.

\subsection{Final Remarks}

The potential of the geometrical methods and symmetry principles described in this work  to establish analogies and connections between different interactions clearly supports their vital contribution for unified field theories.  It is quite consensual in the community that the road to unification and quantum gravity will inevitably lead to a new spacetime paradigm. Whether this unification will imply some convergence between spacetime and physical fields into a single ``physical spacetime-matter' entity' is yet to be seen.  It is worth mentioning at such a level the possibility of the discretization of the underlying geometry of the  spacetime, which is suggested on several grounds, in particular, on the connection of metric-affine geometries and solid state physics systems with defects in their microstructure \cite{Lobo:2014nwa}, or on the thermodynamics of gravitational fields and the entropy of spacetime regions (horizons) in terms of internal microstates. This is linked with  the challenge of making compatible the causal/conformal and metric structures of spacetime \cite{Harko:2018ayt} and  the indeterminacy principle of physical fields. Should the latter be merged with spacetime in a single unified entity, then such a manifold would have to be quantized since it needs to include the indeterminacy principle in its intrinsic geometrical nature.

The unified spacetime-matter manifold conjectured above might have the following ingredients: a conformal symmetry at a fundamental level (possibly inside some unified symmetry group as in SUSY-SUGRA); have more than four spacetime dimensions; include complex numbers (so that the internal symmetries of spinors are embedded and intrinsic to the physical unified manifold); have internal and external physical properties such as energy-momentum, hypermomentum, \mbox{$U(1)$ and $SU(N)$} charges; have electromagnetic  (electric permitivity and magnetic permeability) and thermodynamical properties, and include internal degrees of freedom linked by internal $SU(N)$ symmetries as physical degrees of freedom (in analogy to the microscopic sates of macroscopic systems in statistical physics).

To conclude, we add some remarks.  First, it seems clear that the fundamental symmetry principles and geometrical methods in the gauge approach to gravity lead to quite remarkable predictions about the nature of spacetime and gravity, which might be tested with astrophysical (compact objects \cite{Berti:2015itd}) and cosmological (inflation, late-time evolution \cite{Nojiri:2017ncd}) observations. Second, these methods provide insights on the nature of the gravitational currents which stimulate the research on the fundamental nature of matter and its physical conserved properties. Indeed, understanding the hypermomentum currents from a fundamental point of view might lead to remarkable connections between gravity and hadronic physics. Similar comments apply to the appropriate test-matter properties required to probe post-Riemann geometries, which is needed in order to distinguish experimentally between different classical spacetime paradigms. And finally, these methods can provide robust mathematical frameworks for the search of unified field theories, where spacetime, fermions and bosons might be inextricably linked within a common unified physical framework.

\vspace{6pt}



\authorcontributions{All authors contributed equally to the manuscript. All authors have read and agreed to the published version of the manuscript.}

\funding{FC is funded by the Funda\c{c}\~ao para a Ci\^encia e a Tecnologia (FCT, Portugal) doctoral grant No.PD/BD/128017/2016.
FSNL acknowledges support from the FCT Scientific Employment Stimulus contract with reference
CEECIND/04057/2017.
DRG is funded by the \emph{Atracci\'on de Talento Investigador} programme of the Comunidad de Madrid (Spain) No. 2018-T1/TIC-10431, and acknowledges further support from the Ministerio de Ciencia, Innovaci\'on y Universidades (Spain) project No. PID2019-108485GB-I00/AEI/10.13039/501100011033, the Spanish project No. FIS2017-84440-C2-1-P (MINECO/FEDER, EU), the project PROMETEO/2020/079 (Generalitat Valenciana), and the Edital 006/2018 PRONEX (FAPESQ-PB/CNPQ, Brazil) Grant No. 0015/2019. The authors also acknowledge funding from FCT Projects No. UID/FIS/04434/2020, No. CERN/FIS-PAR/0037/2019 and No. PTDC/FIS-OUT/29048/2017. This article is based upon work from COST Actions CA15117  and CA18108, supported by COST (European Cooperation in Science and Technology).}

\acknowledgments{FC thanks the hospitality of the Department of Theoretical Physics and IPARCOS of the Complutense University of Madrid, where part of this work was carried out.}

\conflictsofinterest{The authors declare no conflict of interest.}




\reftitle{References}





\end{document}